\documentclass[aps,pra,twocolumn,showpacs,superscriptaddress,longbibliography]{revtex4-1}
\usepackage{graphicx}   
\usepackage{dcolumn}    
\usepackage{float}      
\usepackage{amssymb}    
\usepackage{amsmath}
\usepackage{nccmath}    
\usepackage{subfigure}    
\usepackage{braket}
\usepackage{mathrsfs}
\usepackage{dsfont}
\usepackage{makecell}
\usepackage[export]{adjustbox}
\usepackage[toc,page]{appendix}
\newcolumntype{P}[1]{>{\centering\arraybackslash}p{#1}}
\newcolumntype{M}[1]{>{\centering\arraybackslash}m{#1}}

\begin{document}

\title{One- and two-qubit gate infidelities due to motional errors in trapped ions and electrons}

\author{R. Tyler Sutherland}
\email{robert.sutherland@utsa.edu}
\affiliation{Department of Electrical and Computer Engineering, Department of Physics and Astronomy, University of Texas at San Antonio, San Antonio, TX 78249, (USA)}
\author{Qian Yu}
\affiliation{Physics Department, University of California, Berkeley, CA 94720, (USA)}
\affiliation{Challenge Institute for Quantum Computation, University of California, Berkeley, CA 94720, (USA)}

\author{Kristin M. Beck}
\affiliation{Lawrence Livermore National Laboratory, 7000 East Avenue Livermore, CA 94550, (USA)}

\author{Hartmut~H\"affner}%
\affiliation{Physics Department, University of California, Berkeley, CA 94720, (USA)}
\affiliation{Challenge Institute for Quantum Computation, University of California, Berkeley, CA 94720, (USA)}
\affiliation{Computational Research Division, Lawrence Berkeley National Laboratory, Berkeley, CA 94720, (USA)}

\date{\today}

\begin{abstract}
In this work, we derive analytic formulae that determine the effect of error mechanisms on one- and two-qubit gates in trapped ions and electrons. First, we analyze, and derive expressions for, the effect of driving field inhomogeneities on one-qubit gate fidelities. Second, we derive expressions for two-qubit gate errors, including static motional frequency shifts, trap anharmonicities, field inhomogeneities, heating, and motional dephasing. We show that, for small errors, each of our expressions for infidelity converges to its respective numerical simulation; this shows our formulae are sufficient for determining error budgets for high-fidelity gates, obviating numerical simulations in future projects. All of the derivations are general to any internal qubit state, and any \textit{mixed} state of the ion crystal's motion that is diagonal in the Fock state basis. Our treatment of static motional frequency shifts, trap anharmonicities, heating, and motional dephasing apply to both laser-based and laser-free gates, while our treatment of field imhomogenieties applies to laser-free systems.
\end{abstract}
\maketitle

\section{Introduction}
The highest-fidelity quantum computing gates are, at present, performed with trapped ions \cite{srinivas_2021, ballance_2016, gaebler_2016}. This, in combination with long coherence times, inherent uniformity, and all-to-all connectivity, is why trapped ions are one of the most promising quantum computing platforms to date \cite{cirac_1995, monroe_1995, wineland_1998,nielsen_2010, haffner_2008,blatt_2008,harty_2014}. The most common method for performing high-fidelity gates is to couple the internal states of the ions using lasers. While laser-based gates have many advantages, strong spin-motion coupling for example, they suffer from photon scattering and phase-noise. Further, the lasers necessary for high-fidelity gates are expensive and difficult to calibrate. Laser-free gates, however, offer a promising alternative to this paradigm, where laser-fields are replaced with microwaves that directly couple internal states of the ions \cite{mintert_2001,ospelkaus_2008,ospelkaus_2011,harty_2014,harty_2016,srinivas_2018,webb_2018,zarantonello_2019,srinivas_2021}. First, the use of microwave fields eliminates photon scattering. Secondly, the phase and amplitude of microwave fields are easier to control, thereby reducing decoherence due to noisy driving fields, which is often a limiting factor in laser-based gates. However, the relatively slow gate times of microwave gates (compared with laser-based gates) exaggerate the effects of motional decoherence. Trapped electrons, albeit significantly less explored than ions, are another promising qubit platform, and will likely have gate operations similar to those in laser-free trapped ion setups \cite{daniilidis_2013,peng_2017,matthiesen_2021}. Due to their light mass, we expect trapped electrons will operate on much faster time-scales relative to laser-free trapped ion experiments. Unfortunately, because sideband cooling is not possible, trapped electrons will have to operate at much higher temperatures than trapped ions, and, therefore, are likely to be sensitive to motional decoherence as well.

High-fidelity gates are critical for fault tolerant quantum computation, which requires infidelities ranging from $10^{-2}$ to $10^{-4}$ \cite{campbell_2017}, making it important to quantify sources of infidelity. While many error sources are general to trapped ion (and would-be trapped electron) experiments, their effect on gate fidelity is typically calculated numerically; this leads to duplicate computational effort between research groups. Further, it is difficult to determine how gate fidelities scale with various experimental parameters (such as temperature) when working only with numerical simulations. In this work, we aim to ameliorate these issues by deriving analytic formulae for likely sources of motional decoherence in trapped ions and electrons. The formulae we derive make no assumptions about the initial qubit state of the ion, and assume only that the motion is in an incoherent mixed state, diagonal in the Fock state basis. We also assume that the qubit frequency is very different from the motional frequency of all relevant motional modes. We then compare every formula to its respective numerical simulation, showing that the two calculations converge in the high-fidelity limit for each source of infidelity. In short, this work aims to expedite the formulation of error budgets in future experiments, providing analytic formulae where numerical simulations were needed. Moreover, the derivations provide insight into each error mechanism, and show how they scale with relevant experimental parameters. In this work, we focus on static motional frequency shifts, heating, trap anharmonicities, and motional dephasing, which are major sources of infidelity in \textit{all} trapped ion two-qubit gates. We also explore the effects of field inhomogeneities on one- and two-qubit gate fidelities, which are specific to laser-free systems.

The paper is organized as follows. We first describe the theoretical techniques that we use to derive our analytic expressions for infidelity. We then derive errors in single-qubit gates from field inhomogeneities. We then derive errors for two-qubit gates, including static shifts, trap anharmonicities, field inhomogeneitites, motional dephasing, and motional heating. 

\section{Theory}

For the calculations we present in this work, we focus on one- or two-qubit gates in trapped ions and trapped electrons, studying systems with one or two qubits coupled to one or two phonon modes. We represent the qubit subspace as $\ket{\psi(t)}$, and the motion of the crystal as phonon Fock states $\ket{n}$. The density matrix of the initial state is assumed to be $\hat{\rho}(0)=\ket{\psi(0)}\bra{\psi(0)}\otimes \hat{\rho}_{n}(0)$, where $\hat{\rho}_{n}(0)$ is the density matrix of the phonon subspace:
\begin{eqnarray}\label{eq:phonon_dense}
\hat{\rho}_{n}(0) \equiv \sum_{n}\mathcal{P}_{n}\ket{n}\bra{n},
\end{eqnarray}
diagonal in the Fock state basis. Here, $\mathcal{P}_{n}$ is the probability that the phonon subspace begins in the $\ket{n}$ state. For every calculation below, we determine the fidelity by applying the system's time-propagator to a pure state wave function initialized to $\ket{\psi(0)}\ket{n}$, which gives $\mathcal{F}_{n}$, which can then be averaged over $\mathcal{P}_{n}$ to obtain $\mathcal{F}$; this is mathematically equivalent to solving the master equation for an incoherent mixed state, diagonal in the Fock state basis, and tracing over the motional degree-of-freedom to determine the fidelity $\mathcal{F}$. For each calculation, we leave our final answer in terms of the infidelity for a state with an initial phonon number $\mathcal{I}_{n}$, which allows one to straightforwardly determine the ensemble averaged infidelity $\mathcal{I}$, as discussed below.

For each of the calculations, we consider a gate Hamiltonian $\hat{H}_{g}$; when $\hat{H}_{g}$ acts on a system for a gate time $t_{g}$, it results in an `ideal' time-propagator for the gate that we represent with $\hat{U}_{g}(t)$. We note that, unless it would lead to ambiguities, we will drop the time arguments of operators from here on. Under realistic conditions, the actual Hamiltonian will deviate from $\hat{H}_{g}$, producing a value of $\mathcal{F}$ that is less than one. In this manuscript, we only consider high-fidelity gates ($\mathcal{F}\sim 1$), meaning that we can take each individual error source to be small, and assume their resultant infidelities ($\mathcal{I}\equiv 1-\mathcal{F}$) will be additive. If we represent each source of error with $\hat{H}_{e}$, this makes the total Hamiltonian:
\begin{eqnarray}\label{eq:main_ham}
\hat{H}_{t} = \hat{H}_{g} + \hat{H}_{e}.
\end{eqnarray}
which results in a time-propagator for the system acting under $\hat{H}_{t}$, which we represent as $\hat{U}_{t}$. The fidelity $\mathcal{F}_{n}$ for a system with initial phonon number $n$ being acted on by $\hat{U}_{t}$ is given by:
 \begin{eqnarray}\label{eq:initial_fn}
 \mathcal{F}_{n} &=& \sum_{n^{\prime}}|\bra{\psi(0)}\bra{n^{\prime}}\hat{U}_{g}^{\dagger}\hat{U}_{t}\ket{\psi(0)}\ket{n}|^{2}
 \end{eqnarray} 
where $\hat{U}_{g}\ket{\psi(0)}$ is the `ideal' target state.

In this work, we isolate the \textit{small} deviations of $\hat{U}_{t}$ from $\hat{U}_{g}$ by factoring the total time-propagator such that 
\begin{eqnarray}\label{eq:factorize_ut}
\hat{U}_{t} = \hat{U}_{g}\hat{U}_{e}.
\end{eqnarray}
If this factorization is straightforward, we can immediately rewrite Eq.~(\ref{eq:initial_fn}) as:
\begin{eqnarray}\label{eq:fn_just_error}
 \mathcal{F}_{n} &=& \sum_{n^{\prime}}|\bra{\psi(0)}\bra{n^{\prime}}\hat{U}_{e}\ket{\psi(0)}\ket{n}|^{2}.
\end{eqnarray}
We can, subsequently, Taylor expand $\hat{U}_{e}$, and determine the leading-order correction to $\mathcal{F}_{n}$ due to $\hat{H}_{e}$. When it is not straightforward to factor $\hat{U}_{t}$, we transform into the interaction picture with respect to $\hat{H}_{g}$, then use $2^{\text{nd}}$-order time-dependent perturbation theory to approximate $\hat{U}_{e}$. Similar techniques have been used before Ref.~\cite{ball_2014, haddadfarshi_2016}, and date back to work on NMR \cite{haeberlen_1968,ernst_1987}. The interaction picture Hamiltonian is given by:
\begin{eqnarray}
\hat{H}_{I}(t) = \hat{U}_{g}^{\dagger}(t)\hat{H}_{t}(t)\hat{U}_{g}(t) + i\hbar\dot{\hat{U}}_{g}^{\dagger}(t)\hat{U}_{g}(t).
\end{eqnarray}
Upon doing this, we can determine the time-propagator for a system acting under a small $\hat{H}_{I}$ for a time $t_{g}$ by using $2^{\text{nd}}$-order time-dependent perturbation theory:
\begin{eqnarray}\label{eq:time_dep_orig}
\hat{U}_{I}\simeq \hat{I} -\frac{i}{\hbar}\int^{t_{g}}_{0}\!\!dt^{\prime}\hat{H}_{I}(t^{\prime}) - \frac{1}{\hbar^{2}}\int^{t_{g}}_{0}\!\!\!\int^{t^{\prime}}_{0}\!\!dt^{\prime}dt^{\prime\prime}\hat{H}_{I}(t^{\prime})\hat{H}_{I}(t^{\prime\prime}). \nonumber \\
\end{eqnarray}
Transforming out of the interaction picture, this gives:
\begin{eqnarray}
\ket{\psi(t_{g})} = \hat{U}_{g}\hat{U}_{e}\ket{\psi(0)},
\end{eqnarray}
where we have replaced $\hat{U}_{I}$ with $\hat{U}_{e}$, as that the two operators are synonymous in this frame. We have now factorized $\hat{U}_{t}=\hat{U}_{g}\hat{U}_{e}$, at which point Eq.~(\ref{eq:fn_just_error}) applies, and we can determine the leading-order correction to $\mathcal{F}_{n}$ using the expansion given by Eq.~(\ref{eq:time_dep_orig}).

For all of the calculations below, we first calculate the infidelity of a gate for an initial phonon number $\mathcal{I}_{n}\equiv 1-\mathcal{F}_{n}$. Each value of $\mathcal{I}_{n}$'s dependence on the initial qubit state is written in terms of the variance of an operator $\hat{A}$ that acts on the qubit subspace:
\begin{eqnarray}
\lambda_{\hat{A}}^{2} \equiv \braket{\hat{A}^{2}} - \braket{\hat{A}}^{2},
\end{eqnarray}
such that $\hat{A}\in\{\hat{\sigma}_{\alpha},\hat{S}_{\alpha},\hat{S}_{\alpha}^{2}\}$, where $\hat{\sigma}_{\alpha}$ is a Pauli spin operator with eigenstates pointing in the $\alpha$-direction on the Bloch sphere, and $\hat{S}_{\alpha}=\hat{\sigma}_{\alpha,1}+\hat{\sigma}_{\sigma,2}$ is a collective spin operator acting on qubits $1$ and $2$. Physically, $\lambda_{\hat{A}}^{2}$ encapsulates the degree to which the qubit is initialized to an eigenstate of the gate Hamiltonian, meaning that there is less infidelity when an operation affects the qubit(s) less. We use each equation for $\mathcal{I}_{n}$ to determine an ensemble average over an initial mixed state by summing over the probability distribution $\mathcal{P}_{n}$ via:
\begin{eqnarray}
\mathcal{I} = \sum_{n}\mathcal{P}_{n}\mathcal{I}_{n},
\end{eqnarray}
resulting in the replacement $n^{k}$ with its average over $\mathcal{P}_{n}$ $\overline{n^{k}}$ for each equation. Thus, all of the equations derived below are general to any initial qubit state and ensemble average of Fock states. Finally, we note that we are here discussing the value of $\mathcal{I}$ associated with a \textit{single} gate implementation; if the error mechanisms discussed in this work significantly change the motional state, and the motion is not sympathetically cooled between gate operations, our formulae could become less accurate in some cases. 

\section{Single-qubit gate errors}

For the ideal case, we represent a single-qubit gate Hamiltonian with:
\begin{eqnarray}\label{eq:single_ideal}
\hat{H}_{1g} = \hbar\Omega_{1g}\hat{\sigma}_{\alpha}.
\end{eqnarray}
This Hamiltonian is in the rotating frame with respect to the qubit frequency, and we have made the rotating wave approximation with respect to terms oscillating near the qubit frequency. Here, $\Omega_{1g}$ is the Rabi frequency of the microwave field oscillating in the $\hat{n}_{\alpha}$ direction, and $\hat{\sigma}_{\alpha} = (\hat{n}_{\alpha}\cdot \vec{\sigma})$ is a single-qubit Pauli operator, with eigenvectors that point in the $\pm \alpha$ direction on the Bloch sphere. We also assume that the qubit frequency deviates significantly from the motional frequency. After a gate time $t_{g}$, the time-propagator for $\hat{H}_{1g}$ is:
\begin{eqnarray}
\hat{U}_{1g} = e^{-i\Omega_{1g}t_{g}\hat{\sigma}_{\alpha}},
\end{eqnarray}
which (by definition) would give $\mathcal{F}=1$ in the absence of an error term. In this section, we consider the infidelity of gates generated by Eq.~(\ref{eq:single_ideal}) in the presence of inhomogeneities in the microwave field.

\subsection*{Driving field inhomogeneity}

Inhomogeneities of the driving field do not constitute a significant part of the error budget of the highest-fidelity single-qubit laser-free gates in trapped ions \cite{harty_2014}. Due to their light mass, though, trapped electrons typically have a relatively large spatial extent compared to ions. The problem is exacerbated by the absence of laser cooling methods, likely leading to substantially higher motional temperatures. This could render inhomogeneities a significant part of the trapped electron single-qubit error budget. Here, we provide analytical and numerical quantification of the infidelity due to these errors. We represent the error Hamiltonian for driving field inhomogeneities in the presence of two phonon modes as:
\begin{eqnarray}
\tilde{H}_{1e} &=& \hbar\Big\{\omega_{a}\hat{a}^{\dagger}\hat{a} + \omega_{b}\hat{b}^{\dagger}\hat{b} + \hat{\sigma}_{\alpha}\Big(\Omega^{\prime}_{a}[\hat{a}^{\dagger}+\hat{a}] + \Omega^{\prime}_{b}[\hat{b}^{\dagger}+\hat{b}] \nonumber \\
&&+ \Omega_{a}^{\prime\prime}[\hat{a}^{\dagger} + \hat{a}]^{2} + \Omega_{b}^{\prime\prime}[\hat{b}^{\dagger}+\hat{b}]^{2} \nonumber \\
&&+\Omega_{ab}^{\prime\prime}[\hat{a}^{\dagger} + \hat{a}][\hat{b}^{\dagger} + \hat{b}]\Big)\Big\},
\end{eqnarray}
where the tilde indicates that we are in the lab-frame with respect to the motion, $\omega_{a(b)}$ is the motional frequency of the $a(b)$ mode, $\Omega_{a(b)}^{\prime}\equiv \sqrt{\hbar/2m\omega_{a(b)}}\partial\Omega_{1g}/\partial \hat{x}_{a(b)}$ is the Rabi frequency of the $1^{\text{st}}$-derivative of the driving field projected on along the $a(b)$ mode, $\Omega_{a(b)}^{\prime\prime}\equiv (\hbar/4m\omega_{a(b)})\partial^{2}\Omega_{1g}/\partial\hat{x}_{a(b)}^{2}$ is the Rabi frequency of the $2^{\text{nd}}$-derivative of the driving field along the $a(b)$ mode, and $\Omega_{ab}^{\prime\prime}\equiv (\hbar/2m\sqrt{\omega_{a}\omega_{b}})\partial^{2}\Omega_{1g}/\partial\hat{x}_{a}\partial\hat{x}_{b}$ is the cross-Kerr coupling Rabi frequency between modes $a$ and $b$; here, each partial derivative is evaluated at the qubit's location. Moving into the rotating frame with respect to each mode's frequency, and dropping the $2^{\text{nd}}$-order terms that oscillate near $2\omega_{a(b)}$ and $\omega_{a}+\omega_{b}$, we get:
\begin{eqnarray}\label{eq:field_inhomo_1}
\hat{H}_{1e} &\simeq & \hbar\hat{\sigma}_{\alpha}\Big\{\Omega_{a}^{\prime}\Big(\hat{a}^{\dagger}e^{i\omega_{a}t}\! + \hat{a}e^{-i\omega_{a}t}\Big)\! + \Omega_{b}^{\prime}\Big(\hat{b}^{\dagger}e^{i\omega_{b}t}\!+\hat{b}e^{-i\omega_{b}t} \Big) \nonumber \\
&&+ \Omega_{a}^{\prime\prime}\Big(2\hat{a}^{\dagger}\hat{a} + 1\Big) + \Omega_{b}^{\prime\prime}\Big(2\hat{b}^{\dagger}\hat{b} + 1\Big) \nonumber \\
&& + \Omega_{ab}^{\prime\prime}\Big(\hat{a}^{\dagger}\hat{b}e^{i\omega_{ab}t} + \hat{a}\hat{b}^{\dagger}e^{-i\omega_{ab}t}\Big)\Big\},
\end{eqnarray}
where $\omega_{ab}\equiv \omega_{a}-\omega_{b}$. Note that the above equation comprises what are effectively three different error types: $\propto \Omega_{a(b)}^{\prime}$ terms that arise from the $1^{\text{st}}$-derivative of the driving field's projection along the $a(b)$ mode, $\propto \Omega_{a(b)}^{\prime\prime}$ terms that arise from the projection of the $2^{\text{nd}}$-derivative of the driving field on the $a(b)$ mode, and $\propto \Omega_{ab}^{\prime\prime}$ cross-Kerr terms. 

\begin{center}
\begin{figure*}
\includegraphics[width=\textwidth]{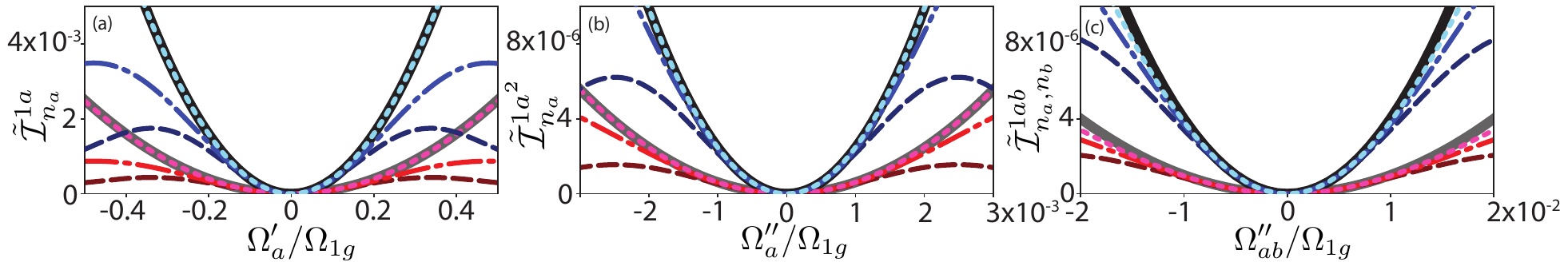}
\caption{Comparison between the one-qubit gate infidelities due to field inhomogeneities given by numerical simulations (dashed lines) and the analytic predictions (solid lines) given in the text. Each graph shows the gate infidelity, normalized by the phonon dependence of the infidelity predicted by the analytic formulae, $\tilde{\mathcal{I}}_{n}$ versus the relative size of the error term compared to the gate Rabi frequency $\Omega_{1g}$. Each gate is implemented for a time $t_{g}=\pi/2\Omega_{1g}$. For each error source, we compare for two initial states: $\ket{\psi(0)}=\ket{\downarrow}$ (analytic: upper black solid, numeric: upper blue), and $\ket{\psi(0)}=\sqrt{3/4}\ket{\downarrow}+\sqrt{1/4}\ket{\uparrow}$ (analytic: lower grey solid, numerical: lower red). (a) Shows $\tilde{\mathcal{I}}^{1a}_{n_{a}}=\mathcal{I}^{1a}_{n_{a}}/(2n_{a}+1)$ versus the first-order field inhomogeneity $\Omega_{a}/\Omega_{1g}$, where $\omega_{a}=10\Omega_{1g}$. This is shown for initial phonon numbers of $n_{a}=0$ (dotted), $n_{a}=100$ (dashed dotted), and $n_{a}=200$ (dashed). (b) Shows $\tilde{\mathcal{I}}^{1a^{2}}_{n_{a}}=\mathcal{I}^{1a^{2}}_{n_{a}}/(2n_{a}+1)^{2}$ versus the strength of the second-order field inhomogeneity $\Omega_{a^{2}}/\Omega_{1g}$, for $n_{a}=0$ (dotted), $n_{a}=100$, and $n_{a}=200$. (c) Shows $\tilde{\mathcal{I}}^{1ab}_{n_{a},n_{b}}=\mathcal{\mathcal{I}}^{1ab}_{n_{a},n_{b}}/(2n_{a}n_{b} + n_{a} + n_{b})$ versus cross-Kerr coupling strength $\Omega_{ab}/\Omega_{1g}$, when $\omega_{ab}=10\Omega_{1g}$.}
\label{fig:coherent_one_qubit}
\end{figure*}
\end{center}

\subsubsection*{$1^{\text{st}}$-order inhomogeneity}

In this subsection, we consider the infidelity $\mathcal{I}^{1a}$ of a single-qubit gate in the presence of a non-zero projection of the field gradient onto mode $a$, noting that the calculation for mode $b$ is identical. This is given by the Hamiltonian:
\begin{eqnarray}
\hat{H}_{1e}^{a} = \hbar\Omega^{\prime}_{a}\hat{\sigma}_{\alpha}\Big\{\hat{a}^{\dagger}e^{i\omega_{a}t} + \hat{a}e^{-i\omega_{a}t} \Big\}.
\end{eqnarray}
The total Hamiltonian for the system is then:
\begin{eqnarray}\label{eq:one_qubit_disp}
\hat{H}_{1t}^{a} &=& \hat{H}_{1g} + \hat{H}_{1e}^{a}.
\end{eqnarray}
In this situation, $\hat{H}_{1e}^{a}$ creates a residual spin-dependent displacement at $t_{g}$, decohering the system.

Because $\hat{H}_{1g}$ and $\hat{H}_{1e}^{a}$ commute at all times, we can immediately factor their time-propagator:
\begin{eqnarray}
\hat{U}_{1t}^{a} = \hat{U}_{1g}\hat{U}_{1e}^{a},
\end{eqnarray}
allowing us to apply Eq.~(\ref{eq:fn_just_error}) to calculate the infidelity of a state initialized to phonon mode $n_{a}$ for this error source $\mathcal{I}^{1a}_{n_{a}}$. We calculate $\hat{U}_{1e}^{a}$ using the Magnus expansion \cite{magnus_1954} for $\hat{H}_{1e}^{a}$. Up to a phase, this is described by a spin-dependent displacement operator:
\begin{eqnarray}
\hat{U}_{1e}^{a} &=& \exp\Big\{-\frac{i}{\hbar}\int^{t_{g}}_{0}dt^{\prime}\hat{H}_{1e}^{a}(t^{\prime}) \Big\} \nonumber \\
&=& \exp\Big\{-\frac{2i\Omega_{a}^{\prime}}{\omega_{a}}\hat{\sigma}_{\alpha}\sin\Big(\frac{\omega_{a}t_{g}}{2}\Big) \nonumber \\
&& \hspace{15mm}\times \Big(\hat{a}^{\dagger}e^{i\omega_{a}t_{g}/2} + \hat{a}e^{-i\omega_{a}t_{g}/2} \Big) \Big\},
\end{eqnarray}
which we can plug into Eq.~(\ref{eq:fn_just_error}). Since we assume $\hat{U}_{1e}^{a}\sim \hat{I}$, we can subsequently insert its Taylor series, and keep only the quadratic contributions to the fidelity $\mathcal{F}_{n_{a}}^{1a}$. This expression for $\mathcal{F}_{n_{a}}^{1a}$, keeping terms up to $\propto (\Omega_{a}^{\prime}/\omega_{a})^{2}$, is:
\begin{eqnarray}
\label{eq:1Q 1st order field inhomogeneity}
\mathcal{F}_{n_{a}}^{1a} & \simeq & 1 - \frac{4\Omega_{a}^{\prime 2}}{\omega_{a}^{2}}\sin^{2}\Big(\frac{\omega_{a}t_{g}}{2}\Big)\Big(2n_{a}+1\Big)\lambda^{2}_{\hat{\sigma}_{\alpha}}.
\end{eqnarray}
where we have simplified this expression by substituting the $t=0$ variance of the $\hat{\sigma}_{\alpha}$ operator: $1 - \braket{\psi(0)|\hat{\sigma}|\psi(0)}^{2}=\braket{\hat{\sigma}_{\alpha}^{2}}-\braket{\hat{\sigma}_{\alpha}}^{2}\equiv \lambda^{2}_{\hat{\sigma}_{\alpha}}$. We can further simplify our expression by taking the time-average of $\sin^{2}(\omega_{a}t_{g}/2)\simeq 1/2$. We here note that if an experiment has sufficient control over $t_{g}$, this step is not necessary and the error can be eliminated by setting $\omega_{a}t_{g}$ to be an integer multiple of $2\pi$. This gives an equation for the infidelity of a gate acting on a state beginning with $n_{a}$ phonons:
\begin{eqnarray}
\label{eq:1Q 1st order field inhomogeneity infidelity}
\mathcal{I}_{n_{a}}^{1a} &\simeq& \frac{2\Omega_{a}^{\prime 2}}{\omega_{a}^{2}}\Big(2n_{a}+1 \Big)\lambda^{2}_{\hat{\sigma}_{\alpha}}.
\end{eqnarray}
In Fig.~\ref{fig:coherent_one_qubit}(a), we compare Eq.~(\ref{eq:1Q 1st order field inhomogeneity infidelity}) to the direct numerical integration of Eq.~(\ref{eq:one_qubit_disp}), varying $\ket{\psi(0)}$ and $n_{a}$, and showing they converge when $|\Omega^{\prime}_{a}/\omega_{a}|\ll 1$.

\subsubsection*{$2^{\text{nd}}$-order inhomogeneity}

If the $2^{\text{nd}}$-derivative of the driving field has a non-zero projection onto phonon mode $a$, noting again the calculation is identical for mode $b$, the Hamiltonian for a single-qubit gate is:
\begin{eqnarray}\label{eq:single_second}
\hat{H}_{1t} &=& \hat{H}_{1g} + \hat{H}_{1e}^{a^{2}} \nonumber \\
&=& \hbar\Omega_{g}\hat{\sigma}_{\alpha} + \hbar\Omega_{a}^{\prime\prime}\hat{\sigma}_{\alpha}\Big(2\hat{a}^{\dagger}\hat{a} + 1 \Big).
\end{eqnarray}
Here, $\hat{H}_{1g}$ and $\hat{H}_{1e}^{a^{2}}$ commute at all times, so we may factor $\hat{U}_{1t}^{a^{2}}=\hat{U}_{1e}^{a^{2}}\hat{U}_{1g}$, allowing us to apply Eq.~(\ref{eq:fn_just_error}), where:
\begin{eqnarray}
\hat{U}_{1e}^{a^{2}} = \exp\Big\{-i\Omega_{a}^{\prime\prime}t_{g}\hat{\sigma}_{\alpha}\Big(2\hat{a}^{\dagger}\hat{a} + 1 \Big) \Big\}.
\end{eqnarray}
Assuming that $|\Omega_{a}^{\prime\prime}t_{g}| \ll 1$, we only keep terms up to $\propto (\Omega_{a}^{\prime\prime}t_{g})^{2}$. The result is:
\begin{eqnarray}
\mathcal{F}_{n_{a}}^{1a^{2}} \simeq 1 - \Omega_{a}^{\prime\prime 2}t_{g}^{2}\Big(2n_{a}+1\Big)^{2}\lambda^{2}_{\hat{\sigma}_{\alpha}}, 
\end{eqnarray}
giving an equation for the infidelity of a gate acting on a state that begins with $n_{a}$ phonons:
\begin{eqnarray}
\label{eq:1Q 2nd order field inhomogeneity infidelity}
\mathcal{I}_{n_{a}}^{1a^{2}}\simeq \Omega_{a}^{\prime\prime 2}t_{g}^{2}\Big(2n_{a}+1\Big)^{2}\lambda^{2}_{\hat{\sigma}_{\alpha}},
\end{eqnarray}
where we have, again, simplified our expression by substituting in the $t=0$ expression for the variance of the $\hat{\sigma}_{\alpha}$ operator. In Fig.~\ref{fig:coherent_one_qubit}(b), we compare Eq.~(\ref{eq:1Q 2nd order field inhomogeneity infidelity}) to the direct numerical simulation of Eq.~(\ref{eq:single_second}), showing they converge when $|\Omega_{a}^{\prime\prime}t_{g}| \ll 1$. 

\subsubsection*{Cross-Kerr coupling}

If the mixed, $2^{\text{nd}}$-order partial derivative of the driving field has a non-zero projection over modes $a$ and $b$, the single-qubit gate is described by:
\begin{eqnarray}\label{eq:cross_kerr_one}
\hat{H}_{1t}^{ab} &=& \hat{H}_{1g} + \hat{H}_{1e}^{ab} \nonumber \\
&=& \hbar\Omega_{1g}\hat{\sigma}_{\alpha} + \hbar\Omega_{ab}^{\prime\prime}\hat{\sigma}_{\alpha}\Big\{\hat{a}^{\dagger}\hat{b}e^{i\omega_{ab}t} + \hat{a}\hat{b}^{\dagger}e^{-i\omega_{ab}t} \Big\}, \nonumber \\
\end{eqnarray}
where $\hat{H}^{ab}_{1e}$ takes the form of a spin-dependent beam-splitter interaction. While $\hat{H}_{1e}^{ab}$ still commutes with $\hat{H}_{1g}$ at all times, allowing us to factor $\hat{U}_{1t}^{ab}$ as $\hat{U}_{1e}^{ab}\hat{U}_{1g}$, the $2^{\text{nd}}$-order term in the Magnus expansion in $\hat{U}_{1e}^{ab}$'s contribution to the dynamics is no longer merely a global phase. Thus, keeping terms up to $2^{\text{nd}}$-order in $\hat{U}_{1e}^{ab}$ gives:
\begin{eqnarray}
\hat{U}_{1e}^{ab}&\simeq& \exp\Big\{-\frac{i}{\hbar}\int^{t_{g}}_{0}dt^{\prime}H_{1e}^{ab}(t^{\prime}) \nonumber \\
&& \hspace{11mm} - \frac{1}{2\hbar^{2}}\int^{t_{g}}_{0}\int^{t^{\prime}}_{0}dt^{\prime}dt^{\prime\prime}\Big[\hat{H}_{1e}^{ab}(t^{\prime}),\hat{H}_{1e}^{ab}(t^{\prime\prime}) \Big] \Big\} \nonumber \\
&=& \exp\Big\{\!\!-i\hat{\sigma}_{\alpha}\Big(\chi\hat{a}^{\dagger}\hat{b} +\!\! \chi^{*}\hat{a}\hat{b}^{\dagger}\Big)\!\!-i\beta\Big(\hat{a}^{\dagger}\hat{a}\! -\!\! \hat{b}^{\dagger}\hat{b} \Big)\Big\},
\end{eqnarray}
where we have defined:
\begin{eqnarray}
\chi &\equiv& \frac{2\Omega^{\prime\prime}_{ab}}{\omega_{ab}}\sin\Big(\frac{\omega_{ab}t_{g}}{2} \Big)e^{i\omega_{ab}t_{g}/2} \nonumber \\
\beta & \equiv & \frac{\Omega_{ab}^{\prime\prime 2}}{\omega_{ab}}\Big\{t_{g} - \frac{\sin(\omega_{ab}t_{g})}{\omega_{ab}} \Big\},
\end{eqnarray}
up to a phase. Since $\hat{H}_{1e}^{ab}$ couples two modes, when applying Eq.~(\ref{eq:fn_just_error}) we must consider both when calculating the fidelity:
\begin{eqnarray}
\!\!\!\!\!\!\mathcal{F}_{n_{a},n_{b}}^{1ab} \!\!=\!\!\!\! \sum_{n^{\prime}_{a},n^{\prime}_{b}}\!\!|\bra{\psi(0)}\bra{n^{\prime}_{a}}\bra{n^{\prime}_{b}}\hat{U}_{1e}^{ab}\ket{\psi(0)}\ket{n_{a}}\ket{n_{b}}|^{2}.
\end{eqnarray}
Applying this equation and, again, keeping only the leading-order contribution to the fidelity $\hat{F}_{n_{a},n_{b}}^{ab}$, we get:
\begin{eqnarray}
\mathcal{F}^{1ab}_{n_{a},n_{b}} &\simeq& 1 - \frac{4\Omega_{ab}^{\prime\prime 2}}{\omega^{2}_{ab}}\sin^{2}\Big(\frac{\omega_{ab}t_{g}}{2}\Big)\Big(2n_{a}n_{b}+n_{a}+n_{b} \Big)\lambda^{2}_{\hat{\sigma}_{\alpha}}, \nonumber \\
\end{eqnarray}
which we can simplify by substituting in the variance of the $\hat{\sigma}_{\alpha}$ operator, and the time-average of the sinusoidal oscillation $\sin^{2}(\omega_{ab}t_{g}/2)\simeq 1/2$. This gives:
\begin{eqnarray}
\label{eq:1Q cross kerr field inhomogeneity infidelity}
\mathcal{I}_{n_{a},n_{b}}^{1ab} \simeq \frac{2\Omega^{\prime\prime 2}_{ab}}{\omega^{2}_{ab}}\Big(2n_{a}n_{b} + n_{a}+n_{b}\Big)\lambda^{2}_{\hat{\sigma}_{\alpha}}.
\end{eqnarray}
In Fig.~\ref{fig:coherent_one_qubit}(c), we compare this result to the direct numerical integration of Eq.~(\ref{eq:cross_kerr_one}), showing that the two calculations converge when $|\Omega_{ab}^{\prime\prime}/\omega_{ab}|\ll 1$.

\begin{table*}[t]
  \centering
\begin{tabular}{ |M{5cm}| M{2cm}| M{6.5cm}|  }
 \hline
 \multicolumn{3}{|c|}{Single-Qubit Gate Errors} \\
 \hline
 \Xhline{2\arrayrulewidth}
 \hline
 error name & equation & infidelity \\
 \hline
 \end{tabular}
 \renewcommand{\arraystretch}{3}
 \begin{tabular}{ |M{5cm}| M{2cm}| M{6.5cm}|  }
 \hline
 1${}^{\text{st}}$-order inhomogeniety & Eq.~(\ref{eq:1Q 1st order field inhomogeneity infidelity}) & $\mathcal{I}^{1a}=\frac{2\Omega_{a}^{\prime 2}}{\omega_{a}^{2}}\Big(2\bar{n}_{a}+1 \Big)\lambda^{2}_{\hat{\sigma}_{\alpha}}$  \\ \hline
2${}^{\text{\text{nd}}}$-order inhomogeniety & Eq.~(\ref{eq:1Q 2nd order field inhomogeneity infidelity}) & $\mathcal{I}^{1a^{2}}=\Omega_{a}^{\prime\prime 2}t_{g}^{2}\Big(4\overline{n_a^2}+4\bar{n}_{a}+1\Big)\lambda^{2}_{\hat{\sigma}_{\alpha}}$ \\ \hline
cross-Kerr coupling & Eq.~(\ref{eq:1Q cross kerr field inhomogeneity infidelity}) & $\mathcal{I}^{1ab}=\frac{2\Omega^{\prime\prime 2}_{ab}}{\omega^{2}_{ab}}\Big(2\bar{n}_{a}\bar{n}_{b} + \bar{n}_{a}+\bar{n}_{b}\Big)\lambda^{2}_{\hat{\sigma}_{\alpha}}$ \\ \hline
\end{tabular}
\caption{Summary of infidelities for single-qubit gates. See text for variable definitions.}
\end{table*}

\section{Two-qubit gate errors}
We represent the, idealized, two-qubit gate Hamiltonian:
\begin{eqnarray}\label{eq:pre_rot_two_qubit}
\tilde{H}_{2g} &=& \frac{\hbar\omega_{0}}{2}\hat{S}_{z} + \hbar\omega_{a}\hat{a}^{\dagger}\hat{a}+ 2\hbar\Omega_{2g}f(t)\hat{S}_{\alpha}\Big(\hat{a}^{\dagger} + \hat{a} \Big), \nonumber \\
\end{eqnarray}
where, again, the tilde indicates that we are working in the lab-frame, $\hat{S}_{\alpha}\equiv \hat{\sigma}_{\alpha,1} + \hat{\sigma}_{\alpha,2}$ is a multi-qubit Pauli spin operator with eigenvectors pointing in the $\alpha$ direction, $\Omega_{2g}$ is the two-qubit gate Rabi frequency, and $f(t)$ is a sinusoidal function representing the temporal dependence of the gradient field, which is either $f(t)\equiv\cos([\omega_{a}-\Delta]t)$ if $\alpha \equiv z$, or $f(t) \equiv \cos([\omega_{0}-\omega_{a} + \Delta]t) + \cos([\omega_{0}+\omega_{a} -\Delta]t)$ if $\alpha$ is in the $xy$-plane; in both bases, $\Delta$ acts as the detuning of the gate. Transforming into the rotating frame with respect to the qubit and motion, as well as making the rotating wave approximation, gives:
\begin{eqnarray}\label{eq:two_qubit_gate}
    \hat{H}_{2g} = \hbar\Omega_{2g}\hat{S}_{\alpha}\Big(\hat{a}^{\dagger}e^{i\Delta t} + \hat{a}e^{-i\Delta t}\Big),
\end{eqnarray}
the form of which can be generated using lasers \cite{molmer_1999,molmer_2000,leibfried_2003} and with microwaves \cite{mintert_2001,ospelkaus_2008,ospelkaus_2011, weidt_2016, harty_2016,sutherland_2019,sutherland_2020,srinivas_2021}. We will begin with Eq.~(\ref{eq:two_qubit_gate}) for all of the following sections except Sec.~\ref{sec:inhomo}.

We analyze the unitary evolution of Eq.~(\ref{eq:two_qubit_gate}) under the influence each subsection's error term, giving a total Hamiltonian $\hat{H}_{2t}$. We continue to assume each error's contribution to the final infidelity will be additive. We analyze the unitary evolution using the Magnus expansion \cite{magnus_1954}, keeping terms up to $2^{\text{nd}}$-order:
\begin{eqnarray}\label{eq:magnus}
    \hat{U}_{2t} &=& \exp\Big(-\frac{i}{\hbar}\int^{t}_{0}dt^{\prime}\hat{H}_{2t}(t^{\prime}) \nonumber \\
    &&-\frac{1}{2\hbar^{2}}\int^{t}_{0}\int^{t^{\prime}}_{0}dt^{\prime}dt^{\prime\prime}[\hat{H}_{2t}(t^{\prime}),\hat{H}_{2t}(t^{\prime\prime})] \Big). 
\end{eqnarray}
For geometric phase gates, the first term in the Magnus expansion represents a spin-dependent displacement operator, making a circular trajectory in phase-space. Ideally, this term will disappear after each of $N$ `loops' in phase-space, occurring every integer multiple of $t=2\pi/\Delta$. Because the spin and motion are entangled during a loop, any decoherence of the motion will affect the spin, and, ultimately, the fidelity of the gate. For an ideal gate, after a time $t_{g} = 2\pi N/\Delta$, this gives:
\begin{eqnarray}
    \hat{U}_{2g} &=& e^{i\phi\hat{S}_{\alpha}^{2}}, 
\end{eqnarray}
where, in this work, we set the phase $\phi = 2\pi N\Omega_{2g}^{2}/\Delta^{2} = \pi/8$, to give a maximally entangled Bell state when operating on $\ket{\psi(0)}=\ket{\downarrow\downarrow}$. This may be turned into an $N$-loop gate by setting $t_{g}\rightarrow N^{1/2}t_{g}$ and $\Delta\rightarrow N^{1/2}\Delta$. Finally, we note that we will frequently use the fact that $\hat{S}_{\alpha}^{2(k+1)} = 4^{k}\hat{S}_{\alpha}^{2}$ and $\hat{S}_{\alpha}^{2k+1}=4^{k}\hat{S}_{\alpha}$, where $k\in\mathbb{N}$.

\subsection{Static motional frequency shifts}\label{sec:static_motion}

During a two-qubit gate, if the frequency of the motional mode is shifted from its idealized value the phase-space trajectory is distorted, giving an error. In this section, we will show that the effects of this error mechanism may be separated into two physical sources: residual spin-motion entanglement from the phase-space trajectory, and the area encompassed in phase-space deviating from its idealized value. These two effects require different techniques to ameliorate their deleterious effects on the gate fidelity.

The total gate Hamiltonian including a static motional frequency shift is:
\begin{eqnarray}\label{eq:static_shift_two_ham}
    \hat{H}_{2t} &=& \hat{H}_{2g}+\hat{H}_{2e}^{\delta} \nonumber \\
    &=&\hbar\Omega_{2g}\hat{S}_{\alpha}\Big(\hat{a}e^{-i\Delta t} + \hat{a}^{\dagger}e^{i\Delta t}\Big) + \hbar\delta\hat{a}^{\dagger}\hat{a},
\end{eqnarray}
where $\delta \ll \Omega_{2g}$ is the frequency shift. Unlike the error mechanisms we have explored thus far, $[\hat{H}_{2g},\hat{H}_{2e}^{\delta}]\neq 0$, meaning we cannot directly factor $\hat{U}_{2t}^{\delta}$. In order to put $\hat{U}_{2t}^{\delta}$ in a form that enables factorization, we transform the above equation into the rotating frame with respect to $\hat{H}_{2e}^{\delta}$, which gives:
\begin{eqnarray}
    \hat{H}_{2g,I} = \hbar\Omega_{2g}\hat{S}_{\alpha}\Big(\hat{a}e^{-i[\Delta +\delta]t} + \hat{a}^{\dagger}e^{i[\Delta + \delta]t} \Big).
\end{eqnarray}
We are now in a frame rotating at $\Delta+\delta$, noting that this will have no effect on $\mathcal{I}^{2\delta}_{n_{a}}$. Plugging this into Eq.~(\ref{eq:magnus}), and dropping terms higher-order than $\propto \delta^{2}$, we can factorize $\hat{U}_{2t}^{\delta}=\hat{U}_{2g}\hat{U}_{2e}^{\delta}$, up to a global ($n_{a}$ dependent) phase, where:
\begin{eqnarray}\label{eq:error_static_mot_shift}
    \hat{U}_{2e}^{\delta} \simeq \exp\Big(-\frac{i\Omega_{2g}\delta t_{g}}{\Delta}\hat{S}_{\alpha}[\hat{a} + \hat{a}^{\dagger}] - \frac{2i\Omega_{2g}^{2}\delta t_{g}}{\Delta^{2}}\hat{S}_{\alpha}^{2}\Big). \nonumber \\
\end{eqnarray}
The first term in this equation is a displacement operator, representing the error from residual spin-motion entanglement, whereas the second term represents the erroneous area encompassed in phase-space, producing an incorrect geometric phase. Due to the fact that $\Delta \propto N^{1/2}$ and $t_{g}\propto N^{1/2}$ for an $N$-loop gate, we can see that increasing $N$ reduces the error due to the incorrect geometric phase, but does not affect the error due to residual spin-motion entanglement. Because of this, more sophisticated pulse sequences such as Walsh modulations \cite{hayes_2012} (see appendix) or polychromatic gates \cite{haddadfarshi_2016,webb_2018,shapira_2018,sutherland_2020} are needed. We note that, in the appendix, we derive the value of $\mathcal{I}^{2\delta}_{n_{a}}$ for gates undergoing such Walsh modulations. We can plug Eq.~(\ref{eq:error_static_mot_shift}) into Eq.~(\ref{eq:fn_just_error}) to obtain $\mathcal{I}_{n_{a}}^{2\delta}$ up to $\propto (\delta/\Omega_{2g})^{2}$:
\begin{eqnarray}\label{eq:infid_static_mot}
    \mathcal{I}_{n_{a}}^{2\delta} &\simeq & \frac{\pi^{2}\delta^{2}}{64\Omega_{2g}^{2}}\Big[(2n_{a}+1)\lambda_{\hat{S}_{\alpha}}^{2} + \lambda_{\hat{S}_{\alpha}^{2}}^{2}/4N\Big],
\end{eqnarray}
where we have encapsulated the dependence of $\mathcal{I}_{n_{a}}^{2\delta}$ on the initial qubit state with the variances of $\hat{S}_{\alpha}$ and $\hat{S}_{\alpha}^{2}$. In Fig.~\ref{fig:coherent_two_qubit}(a), we compare Eq.~(\ref{eq:infid_static_mot}) to the direct numerical integration of Eq.~(\ref{eq:static_shift_two_ham}), showing they converge when $|\delta/\Omega_{2g}|\ll 1$.

\subsection{Trap anharmonicity}\label{sec:trap_anharm}

We now examine the contribution of anharmonicities in the trapping potential to the two-qubit gate infidelity. Considering that $3^{rd}$-order anharmonicities only contain terms that oscillate near $\omega_{a}$, we make the rotating wave approximation and take the leading-order trap anharmonicity to be a quartic $\propto \hat{x}^{4}$ addition to $\hat{H}_{2g}$. Representing $\hat{H}_{2t}$ in terms of ladder operators, and in the rotating frame with respect to $\omega_{a}$, gives:
\begin{eqnarray}\label{eq:trap_an}
\hat{H}_{2t} &=& \hbar\Omega_{2g}\hat{S}_{\alpha}\Big\{\hat{a}^{\dagger}e^{i\Delta t} + \hat{a}e^{-i\Delta t} \Big\} \nonumber \\
&& \hspace{10mm} + \hbar\varepsilon\Big\{\hat{a}^{\dagger}e^{i\omega_{a} t} + \hat{a}e^{-i\omega_{a} t} \Big\}^{4}.
\end{eqnarray}
Upon making the rotating wave approximation, and dropping a global phase, this can be reduced to give:
\begin{eqnarray}\label{eq:trap_an_red}
\hat{H}_{2t} &\simeq & \hbar\Omega_{2g}\hat{S}_{\alpha}\Big\{\hat{a}^{\dagger}e^{i\Delta t} + \hat{a}e^{-i\Delta t} \Big\} + 6\hbar\varepsilon\Big\{\hat{a}^{\dagger}\hat{a} + (\hat{a}^{\dagger}\hat{a})^{2}\Big\}. \nonumber \\
&=& \hat{H}_{2g} + \hat{H}_{2e}^{\varepsilon}.
\end{eqnarray}
We here transform into the interaction picture with respect to $\hat{H}_{2g}$, using $\hat{U}_{2g}$ described by Eq.~(\ref{eq:magnus}), which gives:
\begin{eqnarray}
\hat{H}_{2I}^{\varepsilon} = 6\hbar\varepsilon\Big\{\!(\hat{a}^{\dagger}\!+\!\hat{S}_{\alpha}\gamma^{*})(\hat{a}\!+\!\hat{S}_{\alpha}\gamma) \!+\! [(\hat{a}^{\dagger}\!+\!\hat{S}_{\alpha}\gamma^{*})(\hat{a}\!+\!\hat{S}_{\alpha}\gamma)]^{2}\Big\}, \nonumber \\
\end{eqnarray}
where:
\begin{eqnarray}\label{eq:gate_gamma}
\gamma \equiv \frac{\Omega_{2g}}{\Delta}\Big\{1-e^{i\Delta t} \Big\}.
\end{eqnarray}
After this transformation, the time propagator for an ideal gate is $\hat{U}_{2I}^{\varepsilon}=\hat{I}$. Since we assume that infidelities are small, $\hat{U}_{2I}^{\varepsilon}\sim \hat{I}$, which we can evaluate using Eq.~(\ref{eq:time_dep_orig}). Before evaluating Eq.~(\ref{eq:time_dep_orig}) we transform out of the interaction picture, giving a factored $\hat{U}_{2t}^{\varepsilon}=\hat{U}_{2g}\hat{U}_{2I}^{\varepsilon}$, and then apply Eq.~(\ref{eq:fn_just_error}). Here, we only wish to extract the leading-order ($\propto [\varepsilon/\Omega_{2g}]^{2}$) contribution to the infidelity. This means that in evaluating the fidelity we can neglect imaginary, and off-diagonal, terms from the $2^{\text{nd}}$-order integral, simplifying the evaluation. Doing this, and then plugging Eq.~(\ref{eq:time_dep_orig}) into Eq.~(\ref{eq:fn_just_error}) gives an infidelity of:
\begin{eqnarray}
\label{eq:2q trap anharmonicity infidelity}
\mathcal{I}_{n_{a}}^{2\varepsilon} &\simeq& \frac{9\pi^{2}\varepsilon^{2}}{16\Omega_{2g}^{2}}\Big\{\lambda^{2}_{\hat{S}_{\alpha}}\Big[4(2 n^{3}_{a}+3n_{a}^{2}+3n_{a}+1) \nonumber \\
&&+ \frac{6}{N}(2n^{2}_{a}+2n_{a}+1) +\frac{9}{4N^{2}}(2n_{a}+1) \Big] \nonumber \\ 
&& + \lambda^{2}_{\hat{S}_{\alpha}^{2}}\Big[\frac{3}{8N}(11n_{a}^{2}+11n_{a}+3) \nonumber \\
&& + \frac{3}{4N^{2}}(2n_{a}+1) +\frac{9}{64N^{3}}\Big]\Big\},
\end{eqnarray}
where we have,  again, encapsulated the dependence on the initial spin with the variances $\lambda_{\hat{S}_{\alpha},\hat{S}_{\alpha}^{2}}$. Figure~\ref{fig:coherent_two_qubit_markovian}(b) compares Eq.~(\ref{eq:2q trap anharmonicity infidelity}) to the direct numerical integration of Eq.~(\ref{eq:trap_an_red}) for various initial states of the qubits and motion, showing the two calculations converge when $|\varepsilon/\Omega_{2g}|\ll 1$.

\subsection{Field inhomogeneities}\label{sec:inhomo}

\begin{center}
\begin{figure*}
\includegraphics[width=\textwidth]{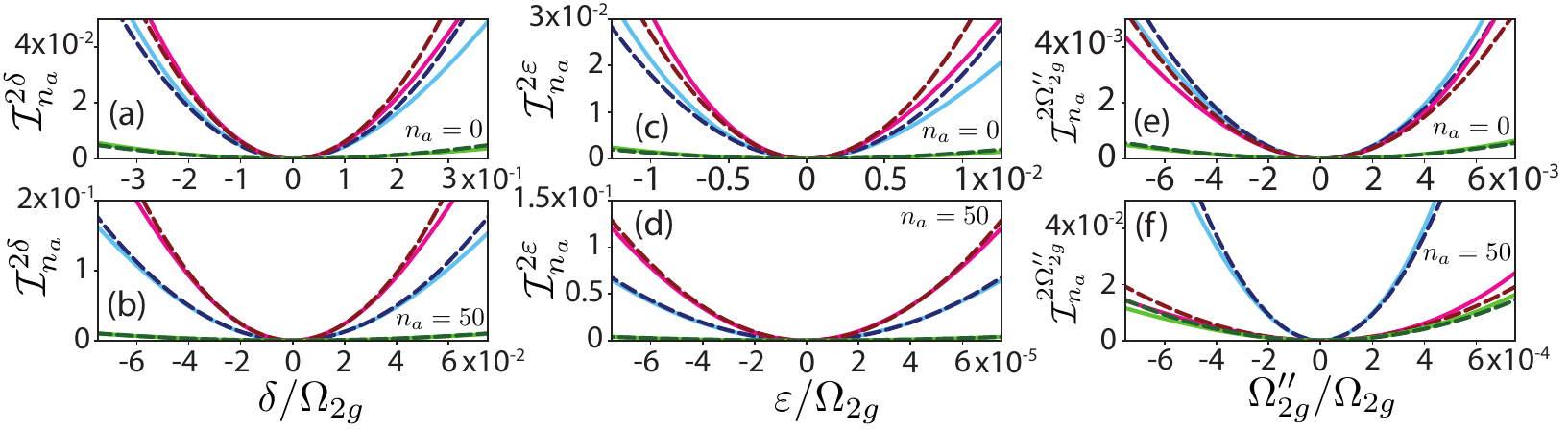}
\caption{Comparison of infidelities $\mathcal{I}_{n_{a}}$ for an initial phonon number $n_{a}$ versus error size between the analytic formulas (solid lines) described in this work and their corresponding numerical simulations (dashed lines). This is shown for initial states: $\ket{\psi(0)}=\ket{\downarrow\downarrow}$ ((a-d) middle (e-f) top blue), $\ket{\psi(0)}=\sqrt{1/3}\ket{\downarrow\downarrow}+\sqrt{2/3}\ket{\uparrow\uparrow}$ ((a-d) top (e-f) middle red), and $\ket{\psi(0)} = \sqrt{1/3}\ket{\downarrow\downarrow}-\sqrt{2/3}\ket{\uparrow\uparrow}$ (bottom green). The left column (a-b) is for a static motional frequency shift $\delta$, the middle column (c-d) is for a trap anharmonicities $\varepsilon_{a}$, and the right column (e-f) is for inhomogeneities of the gradient field $\Omega_{2g}^{\prime\prime}$. The first row (a,c,e) is for ion crystals that begin in the ground state of the phonon mode $n=0$, and the second row (b,d,f) is for ion crystals that begin with a phonon mode such that $n=50$. Note that, for every figure, the values of $\mathcal{I}_{n}$ predicted by the numeric and analytic results converge, for high-fidelity gates.}
\label{fig:coherent_two_qubit}
\end{figure*}
\end{center}

Another error occurs if the gradient field that generates spin-motion coupling for the two-qubit gate changes significantly over the spatial extent of the qubits' motion. In this section, we consider a gradient with non-zero $1^{\text{st}}$ and $2^{\text{nd}}$ derivatives. We begin by transforming Eq.~(\ref{eq:pre_rot_two_qubit}) into the rotating frame with respect to the qubit and motional frequencies, and eliminating terms that oscillate $\propto \omega_{0}$. This gives:
\begin{eqnarray}\label{eq:inhomo_pre}
\hat{H}_{2t} &=& 2\hbar\cos([\omega_{a}-\Delta]t)\hat{S}_{\alpha}\Big\{\Omega_{2g} + \Omega_{2g}^{\prime}\Big(\hat{a}^{\dagger}e^{i\omega_{a}t} + \hat{a}e^{-i\omega_{a}t}\Big)\nonumber \\
&&+ \Omega_{2g}^{\prime\prime}\Big(\hat{a}^{\dagger}e^{i\omega_{a}t} + \hat{a}e^{-i\omega_{a}t}\Big)^{2}\Big\}\Big\{\hat{a}^{\dagger}e^{i\omega_{a}t}+\hat{a}e^{-i\omega_{a}t}\Big\}, \nonumber \\
\end{eqnarray}
where $\Omega_{2g}^{\prime}\equiv (\hbar/2m\omega_{a})^{1/2}\partial\Omega_{2g}/\partial \hat{x}_{a}$, and $\Omega_{2g}^{\prime\prime}\equiv (\hbar/4m\omega_{a})\partial^{2}\Omega_{2g}/\partial \hat{x}_{a}^{2}$. In our treatment, we assume the $1^{\text{st}}-$order $\propto \Omega_{2g}^{\prime}$ term is negligible relative to the $2^{\text{nd}}$-order $\propto \Omega_{2g}^{\prime\prime}$ term, because the former contains only terms that rotate $\propto \omega_{a}$ and the latter contains terms that rotate $\propto \Delta$. This makes the relative contribution of the $\propto \Omega_{2g}^{\prime}$ terms to the $\Omega_{2g}^{\prime\prime}$ terms $\sim(\Omega_{2g}^{\prime}\Delta/\Omega_{2g}^{\prime\prime}\omega_{a})^{2}$; since the value of $(\Delta/\omega_{a})^{2}$ will be very small for both trapped ions and trapped electrons, $\Omega_{2g}^{\prime}$ would need to be several orders-of-magnitude larger than $\Omega_{2g}^{\prime\prime}$ to significantly contribute. Further, because traps are typically designed so that the ions are located near where $\Omega_{2g}$ is at a maximum, they tend to minimize $\Omega_{2g}^{\prime}$ by default. Neglecting the $1^{\text{st}}$-order inhomogeneity, and dropping all terms that oscillate $\propto \omega_{a}$, gives:
\begin{eqnarray}\label{eq:inhomo_two_qubit_ham}
\hat{H}_{2t} &=& \hat{H}_{2g} + 3\hbar\Omega_{2g}^{\prime\prime}\hat{S}_{\alpha}\Big\{\hat{a}^{\dagger}\hat{a}\hat{a}^{\dagger}e^{i\Delta t} +  \hat{a}\hat{a}^{\dagger}\hat{a}e^{-i\Delta t}\Big\}. \nonumber \\
\end{eqnarray}
To calculate the fidelity, we, again, transform into the interaction picture with respect to the ideal gate Hamiltonian $\hat{H}_{2g}$. Transforming into the interaction picture using $\hat{U}_{2g}$ gives:
\begin{eqnarray}
\hat{H}_{2I}^{\Omega_{2g}^{\prime\prime}} &=& 3\hbar\Omega_{2g}^{\prime\prime}\hat{S}_{\alpha}[\hat{a}^{\dagger} + \hat{S}_{\alpha}\gamma^{*}][\hat{a} +\hat{S}_{\alpha}\gamma][\hat{a}^{\dagger} +\hat{S}_{\alpha}\gamma^{*}]e^{i\Delta t} + c.c. \nonumber \\
\end{eqnarray}
After this transformation, we can apply Eq.~(\ref{eq:time_dep_orig}), which results in a factorized $\hat{U}_{2t}^{\Omega_{2g}^{\prime\prime}} = \hat{U}_{2g}\hat{U}_{2I}^{\Omega_{2g}^{\prime\prime}}$ upon transforming out of the interaction picture. We can subsequently apply Eq.~(\ref{eq:fn_just_error}),  and keep only the $(\propto\Omega_{2g}^{\prime\prime}/\Omega_{2g})^{2}$ contributions to the gate fidelity:
\begin{eqnarray}
\label{eq:2Q field inhomogeneities infidelity} 
\mathcal{I}_{n_{a}}^{2\Omega_{2g}^{\prime\prime}} &=& \frac{9\pi^{2}\Omega_{2g}^{\prime\prime 2}}{16\Omega_{2g}^{2}}\Big\{\lambda^{2}_{\hat{S}^{2}_{\alpha}}\Big[4n^{2}_{a} + 4n_{a} + 1+ \frac{3}{2N}\Big(2n_{a}+1\Big) \nonumber \\
&& +\frac{9}{16N^{2}}\Big] + \frac{4}{N}\lambda_{\hat{S}_{\alpha}}^{2}\Big(2n_{a}+1\Big)\Big\},
\end{eqnarray}
giving the infidelity of a two-qubit gate initialized to the motional state $n_{a}$, as well as an initial qubit state with variances $\lambda_{\hat{S}_{\alpha}}^{2}$ and $\lambda_{\hat{S}_{\alpha}^{2}}^{2}$. In Fig.~\ref{fig:coherent_two_qubit}(c), we compare Eq.~(\ref{eq:2Q field inhomogeneities infidelity}) to the direct numerical integration of Eq.~(\ref{eq:inhomo_two_qubit_ham}), showing for various initial states of the qubits and motion that the two calculations converge when $|\Omega^{\prime\prime}_{2g}/\Omega_{2g}|\ll 1$.

\subsection{Infidelities from Markovian bath}\label{sec:markovian}
There is a temporal window during a geometric phase gate where the qubits are entangled to their motion. During this window, the fidelity of the gate is sensitive to extraneous fields that couple to the motion. In this section, we focus on fields that have very low coherence times relative to $t_{g}$, namely those causing heating and motional dephasing. In this work, we eschew the application of the Lindblad equation, typically used to model these two decoherence mechanisms \cite{mandel_1995}. We model our `coupling to a bath' by first calculating the fidelity of a two-qubit gate in the presence of an error of the form:
\begin{eqnarray}\label{eq:markov_ham}
\hat{H}_{2\omega}\propto \cos(\omega t)\hat{M},
\end{eqnarray}
where $\hat{M}$ is an operator describing bath coupling. Subsequently, we average over the normalized power spectral density $S_{\omega}$ of each $\omega$:
\begin{eqnarray}\label{eq:avg_over_omega}
\mathcal{F} = \int^{\infty}_{0}d\omega S_{\omega}\mathcal{F}_{\omega},
\end{eqnarray}
where $\mathcal{F}_{w}$ is the fidelity of a gate undergoing the $\hat{H}_{2\omega}$ perturbation. 

We first show that, upon invoking the Markovian approximation, this prescription is equivalent to the Linblad formalism. Take $\ket{\psi_{\omega}(t)}$ to be a wave function undergoing $\hat{H}_{2t}$. We wish to calculate $\ket{\psi_{\omega}(t+\delta t)}$, then, upon averaging over $S_{\omega}$, show that $\hat{\rho}(t) \equiv \int^{\infty}_{0}d\omega S_{\omega}\hat{\rho}_{\omega}$ follows the Lindblad formalism, where $\hat{\rho}_{\omega}\equiv \ket{\psi_{\omega}(t)}\bra{\psi_{\omega}(t)}$. Assuming $1/\delta t$ is significantly larger than any Rabi frequency in $\hat{H}_{2t}\equiv\hat{H}_{2g}+\hat{H}_{2\omega}$, we can use $2^{\text{nd}}$-order time-dependent perturbation theory to calculate $\ket{\psi_{\omega}(t)}$:
\begin{eqnarray}
\ket{\psi_{\omega}(t\!+\!\delta t)}\! &\simeq&\! \ket{\psi_{\omega}(t)} - \frac{i}{\hbar}\int^{t+\delta t}_{t}dt^{\prime}\hat{H}_{2t}(t^{\prime})\ket{\psi_{\omega}(t)} \nonumber \\
&&\!\!\!- \frac{1}{\hbar^{2}}\!\!\!\int^{t+\delta t}_{t}\!\!\!\!\!\int^{t^{\prime}}_{t}\!\!\!\!\!dt^{\prime}\!dt^{\prime\prime}\!\hat{H}_{2t}(t^{\prime\!})\!\hat{H}_{2t}(t^{\prime\prime}\!)\!\ket{\psi_{\omega}(t)}. \nonumber \\
\end{eqnarray}
We now calculate $\hat{\rho}_{\omega}(t+\delta t)=\ket{\psi_{\omega}(t+\delta t)}\bra{\psi_{\omega}(t+\delta t)}$, while keeping only terms that are linear in $\delta t$, and dropping every term that averages to zero upon integrating over $S_{\omega}$, i.e. terms that are proportional to $\hat{H}_{2\omega}$, $\hat{H}_{2g}\hat{H}_{2\omega}$, or $\hat{H}_{2\omega}\hat{H}_{2g}$. We also replace $\hat{\rho}_{\omega}(t)\rightarrow \hat{\rho}(t)$, equivalent to making the Markovian approximation. This gives:
\begin{eqnarray}\label{eq:lindblad}
\dot{\hat{\rho}}(t) &\simeq& -\frac{i}{\hbar}[\hat{H}_{2g}(t),\hat{\rho}(t)] \nonumber  \\ 
&&+\frac{1}{\hbar^{2}\delta t}\int^{\infty}_{0}\!\!\int^{t+\delta t}_{t}\!\!\!\int^{t+\delta t}_{t}\!\!d\omega dt^{\prime}dt^{\prime\prime}S_{\omega}\hat{H}_{2\omega}(t^{\prime})\hat{\rho}(t)\hat{H}_{2\omega}(t^{\prime\prime}) \nonumber \\
&&-\frac{1}{\hbar^{2}\delta t}\int^{\infty}_{0}\!\!\int^{t +\delta t}_{t}\!\!\!\int^{t^{\prime}}_{t}\!d\omega dt^{\prime}dt^{\prime\prime}S_{\omega}\nonumber \\
&&\times \Big[\hat{H}_{2\omega}(t^{\prime})\hat{H}_{2\omega}(t^{\prime\prime})\hat{\rho}(t)
+ \hat{\rho}(t)\hat{H}_{2\omega}(t^{\prime\prime})\hat{H}_{2\omega}(t^{\prime})\Big],
\end{eqnarray}
taking the form of the Lindblad master equation. Note that in the final line of the equation, we have made the substitution $\dot{\hat{\rho}}(t)\simeq [\hat{\rho}(t+\delta t)-\hat{\rho}(t)]/\delta t $. In the appendix, we show that evaluating the integrals gives the standard form of the master equation for both heating and motional dephasing. 

\begin{center}
\begin{figure*}
\includegraphics[width=\textwidth]{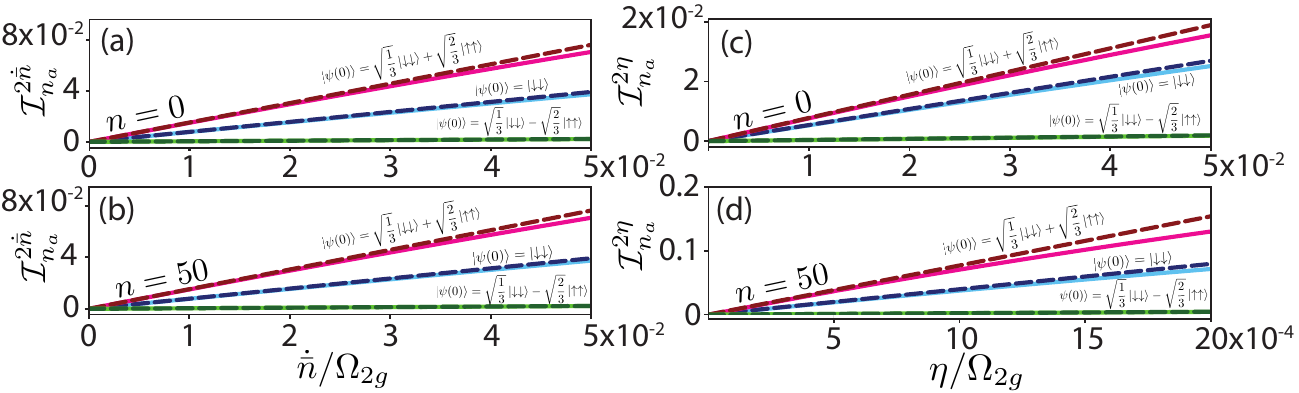}
\caption{Comparison between infidelities for a given initial phonon number $\mathcal{I}_{n}$ between numerical simulations (dashed lines) and the analytic formulas (solid lines) for motional heating and dephasing, as described in this work, versus the bath coupling rate. This is shown for initial states: $\ket{\psi(0)}=\ket{\downarrow\downarrow}$ (middle blue), $\ket{\psi(0)}=\sqrt{1/3}\ket{\downarrow\downarrow}+\sqrt{2/3}\ket{\uparrow\uparrow}$ (top red), and $\ket{\psi(0)} = \sqrt{1/3}\ket{\downarrow\downarrow}-\sqrt{2/3}\ket{\uparrow\uparrow}$ (bottom green). The left column (a-b) motional heating $\dot{\bar{n}}$ and the right column (c-d) is for motional dephasing $\eta$. The first row is for ion crystals that begin in the ground state of the phonon mode $n=0$, and the second row (b,d,f) is for ion crystals that begin with a phonon mode where $n=50$. Note that, for every figure, the values of $\mathcal{I}_{n}$ predicted by the numeric and analytic results converge, for high-fidelity gates.}
\label{fig:coherent_two_qubit_markovian}
\end{figure*}
\end{center}

\subsubsection{Heating}\label{sec:heating}

When an ion crystal is close to a surface, moving charges within the surface create extraneous electric fields that decohere the motion of the crystal \cite{brownnutt_2015}. We assume these fields are homogeneous over the extent of the qubits' motion, and model $\hat{H}_{2\omega}^{g_{h}}$ as:
\begin{eqnarray}\label{eq:heating_no_rot}
\hat{H}_{2\omega}^{g_{h}}(t) &=& \hbar F\cos(\omega t)\hat{x}_{a} \nonumber \\
&=& 2\hbar g_{h}\cos(\omega t)(\hat{a}^{\dagger} + \hat{a}),
\end{eqnarray}
where $F$ is the projection of the electric force onto the designated mode of motion, $\hat{x}_{a}$ is the position operator, and $g_{h}\equiv \frac{1}{2}F\sqrt{\hbar/2m\omega_{a}}$. Writing this in the interaction picture with respect to the frequency of the motional mode $\omega_{a}$, and making the rotating wave approximation, we get:
\begin{eqnarray}\label{eq:interaction_heat}
\hat{H}^{g_{h}\prime}_{2\omega}(t) \simeq \hbar g_{h}(\hat{a}^{\dagger}e^{i(\omega_{a}-\omega)t} + \hat{a}e^{-i(\omega_{a}-\omega)t}).
\end{eqnarray}
We analyze the effect of $\hat{H}^{g_{h}\prime}_{2\omega}$ on $\mathcal{F}_{n_{a},\omega}$ by transforming $\hat{H}_{2t} = \hat{H}_{2g} + \hat{H}^{g_{h}\prime}_{2\omega}$ into the interaction picture with respect to $\hat{H}^{g_{h}\prime}_{2\omega}$. We do this using the transformation:
\begin{eqnarray}
\hat{U}_{\omega} = \exp\Big\{\gamma_{\omega}(t)\hat{a}^{\dagger} - \gamma_{\omega}^{*}(t)\hat{a} \Big\},
\end{eqnarray}
giving a displacement operator that transforms the system into a frame that follows the changes to the `classical' position and momentum induced by the electric field \cite{ehrenfest_1927,heller_1975,sutherland_2021}, up to a phase, where:
\begin{eqnarray}
\gamma_{\omega}(t) = \frac{g_{h}}{\omega - \omega_{a}}\Big[e^{i(\omega_{a}-\omega)t} -1\Big],
\end{eqnarray}
which makes:
\begin{eqnarray}
\hat{H}_{2t}^{g_{h}} &=& \hat{H}_{2g} + \hat{H}_{2e}^{g_{h}} \nonumber \\
&=& \hbar\Omega_{2g}\hat{S}_{\alpha}\Big(\hat{a}^{\dagger}e^{i\Delta t} + \hat{a}e^{-i\Delta t} \Big) \nonumber \\
&&+ \hbar\Omega_{2g}\hat{S}_{\alpha}\Big(\gamma_{\omega}^{*} e^{i\Delta t} + \gamma_{\omega} e^{-i\Delta t}\Big). 
\end{eqnarray}
This equation shows that heating can be represented as a $\propto \hat{S}_{\alpha}$ shift of the qubit in this frame. This is because the electric field displaces the charged particles along the the spin-dependent gradient that is driving the gate, shifting the spin coupling strength. As we will show, this correspondence results in an infidelity that does not depend on the initial state of the motion, and is, therefore, independent of temperature. Noting that $\hat{H}_{2g}$ and $\hat{H}_{2e}^{g_{h}}$ commute at all times, we can factor the time propagator for the whole system $\hat{U}_{2t} = \hat{U}_{2g}\hat{U}_{2e}^{g_{h}}$, where:
\begin{eqnarray}
\hat{U}_{2e}^{g_{h}} = \exp\Big(-i\Omega_{2g}\hat{S}_{\alpha}\int^{t_{g}}_{0}dt^{\prime}\big[\gamma_{\omega}^{*}(t^{\prime})e^{i\Delta t^{\prime}} + \gamma_{\omega}(t^{\prime})e^{-i\Delta t^{\prime}} \big] \Big). \nonumber \\
\end{eqnarray}
Since we assume small errors, we can Taylor expand $\hat{U}_{2e}^{g_{h}}$, and apply Eq.~(\ref{eq:fn_just_error}), which, to leading-order, gives:
\begin{eqnarray}
\mathcal{F}_{n_{a},\omega}^{g_{h}} = 1 - \xi_{\omega}(t_{g})\lambda_{\hat{S}_{\alpha}}^{2},
\end{eqnarray}
where:
\begin{eqnarray}
\xi_{\omega}(t_{g}) \equiv \Omega_{2g}^{2}\int^{t_{g}}_{0}\int^{t_{g}}_{0}dt^{\prime}dt^{\prime\prime} & \Big[\gamma_{\omega}^{*}(t^{\prime})e^{i\Delta t^{\prime}} + \gamma_{\omega}(t^{\prime})e^{-i\Delta t^{\prime}} \Big] \nonumber \\
\times & \Big[\gamma_{\omega}^{*}(t^{\prime\prime})e^{i\Delta t^{\prime\prime}} + \gamma_{\omega}(t^{\prime\prime})e^{-i\Delta t^{\prime\prime}} \Big]. \nonumber \\
\end{eqnarray}
We can now apply Eq.~(\ref{eq:avg_over_omega}) and average over $S_{\omega}$:
\begin{eqnarray}
\mathcal{F}_{n_{a}}^{g_{d}} &= & 1 - \lambda_{\hat{S}_{\alpha}}^{2}\int^{\infty}_{0}d\omega S_{\omega}\xi_{\omega}(t_{g}), 
\end{eqnarray}
leaving a final integral over $\omega$ to determine the fidelity:
\begin{eqnarray}
\xi(t_{g}) \equiv \!\Omega_{2g}^{2}\!\! \int^{\infty}_{0}\!\!\!\!\int^{t_{g}}_{0}\!\!\!\int^{t_{g}}_{0}&\!\!\!\! d\omega dt^{\prime}dt^{\prime\prime}\! S_{\omega}\!\Big[\gamma_{\omega}^{*}(t^{\prime})e^{i\Delta t^{\prime}} \!\!\!\!+\! \gamma_{\omega}(t^{\prime})e^{-i\Delta t^{\prime}} \Big] \nonumber \\
&\!\!\!\!\!\!\!\!\!\!\!\!\!\!\!\!\!\! \times\!\Big[\gamma_{\omega}^{*}(t^{\prime\prime})e^{i\Delta t^{\prime\prime}}\!\! +\! \gamma_{\omega}(t^{\prime\prime})e^{-i\Delta t^{\prime\prime}} \Big].
\end{eqnarray}
Keeping only terms that do not average to zero upon integrating over time (since $t_{g}=2\pi N/\Delta$):
\begin{eqnarray}
\xi(t_{g}) &=& \medint{\int^{\infty}_{0}}\medint{\int^{t_{g}}_{0}}\medint{\int^{t_{g}}_{0}}d\omega dt^{\prime}dt^{\prime\prime}\frac{4g_{h}^{2}S_{\omega}\Omega_{2g}^{2}}{(\omega-\omega_{a})^{2}}\cos(\omega^{\prime}t^{\prime})\cos(\omega^{\prime} t^{\prime\prime}) \nonumber \\
 &= & \medint{\int^{\infty}_{0}}\medint{\int^{t_{g}}_{0}}\medint{\int^{t_{g}}_{-t_{g}}}d\omega dt^{\prime}dt^{\prime\prime}\frac{2g_{h}^{2}S_{\omega}\Omega_{2g}^{2}}{(\omega-\omega_{a})^{2}}\cos(\omega^{\prime}t^{\prime})\cos(\omega^{\prime}t^{\prime\prime}),\nonumber \\
\end{eqnarray}
where $\omega^{\prime}\equiv \omega_{a}-\omega - \Delta$. We now make two approximations: extending the limits of the the first integral from $[t_{g},-t_{g}]$ to $[-\infty,\infty]$, and setting $S_{\omega_{a}-\Delta}\simeq S_{\omega_{a}}$. After this, the above integrals may be straightforwardly integrated, giving:
\begin{eqnarray}
\xi(t_g) &=& \frac{4\pi g_{h}^{2}S_{\omega_{a}}\Omega_{2g}^{2}t_{g}}{\Delta^{2}},
\end{eqnarray}
resulting in a final infidelity of:
\begin{eqnarray}\label{eq:infid_heating_pre}
\mathcal{I}_{n_{a}}^{2g_{h}} = \frac{\pi^{2}g_{h}^{2}S_{\omega_{a}}}{8\Omega_{2g}N^{1/2}}\lambda_{\hat{S}_{\alpha}}^{2},
\end{eqnarray}
We substitute the heating rate $\dot{\bar{n}} = \pi g_{h}^{2}S_{\omega_{a}}$ (see appendix), giving a final form:
\begin{eqnarray}\label{eq:heating_fid}
\mathcal{I}_{n_{a}}^{\dot{\bar{n}}_{a}} = \frac{\pi \dot{\bar{n}}\lambda^{2}_{\hat{S}_{\alpha}}}{8\Omega_{2g}N^{1/2}}.
\end{eqnarray}
Since $\mathcal{I}_{n_{a}}^{\dot{\bar{n}}_{a}}$ is independent of $n_{a}$, as expected, averaging over $\mathcal{P}_{n_{a}}$ simply gives $\mathcal{I}^{\dot{\bar{n}}_{a}}_{n_{a}}=\mathcal{I}^{\dot{\bar{n}}_{a}}$. In Fig.~\ref{fig:coherent_two_qubit_markovian}(a), we compare Eq.~(\ref{eq:heating_fid}) to the direct integration of the heating master equation (see appendix) for various initial states of the qubits and motion, showing that the two converge when $\dot{\bar{n}}/\Omega_{2g}\ll 1$. 

\begin{table*}[t]
\centering
 \begin{tabular}{ |M{3.5cm}| M{1.5cm}| M{11.5cm}|  }
 \hline
 \multicolumn{3}{|c|}{Two-Qubit Gate Errors} \\
 \hline
 \Xhline{2\arrayrulewidth}
 \hline
 error name & equation & infidelity \\
 \hline
 \end{tabular}
 \renewcommand{\arraystretch}{3}
 \begin{tabular}{ |M{3.5cm}| M{1.5cm}| M{11.5cm}|  }
 \hline
static motional shift &Eq.~(\ref{eq:infid_static_mot})& $\mathcal{I}^{2\delta}=\frac{\pi^2\delta^{2}}{64\Omega_{2g}^2}\Big[(2\bar{n}_{a}+1)\lambda_{\hat{S}_\alpha}^2 + \lambda_{\hat{S}_\alpha^2}^2/4N\Big]$ \\
\hline
trap anharmonicity & Eq.~(\ref{eq:2q trap anharmonicity infidelity}) &  $\mathcal{I}^{2\varepsilon}=\frac{9\pi^2\varepsilon^2}{16\Omega_{2g}^2}\Big\{\lambda^2_{\hat{S}_\alpha}\Big[4\Big(2 \overline{n_{a}^3}+3\overline{n_{a}^2}+3\overline{n}_{a}+1\Big) + \frac{6}{N}\Big(2\overline{n_{a}^2}+2\bar{n}_{a}+1\Big) + \frac{9}{4N^2}\Big(2\bar{n}_{a}+1\Big) \Big]$ \\ 
 & &\hspace{35mm} $+ \lambda^2_{\hat{S}_\alpha^2}\Big[\frac{3}{8N}(11\overline{n_{a}^2}+11\bar{n}_{a}+3) + \frac{3}{4N^2}(2\bar{n}_{a}+1) + \frac{9}{64N^3}\Big]\Big\}$  \\
 \hline
field inhomogeneities & Eq.~(\ref{eq:2Q field inhomogeneities infidelity}) &   
$\mathcal{I}^{2\Omega_{2g}^{\prime\prime}}=\frac{9\pi^2\Omega_{2g}^{\prime\prime 2}}{16\Omega_{2g}^2}\Big\{\lambda^2_{\hat{S}^2_\alpha}\Big[4\overline{n_{a}^2} + 4\bar{n}_{a} + 1+ \frac{3}{2N}\Big(2\bar{n}_{a}+1\Big) +\frac{9}{16N^2}\Big] + \frac{4}{N}\lambda_{\hat{S}_\alpha}^2\Big(2\bar{n}_{a}+1\Big)\Big\}$\\
 \hline
heating & Eq.~(\ref{eq:heating_fid}) &  $\mathcal{I}^{2\dot{\bar{n}}_{a}}=\frac{\pi\dot{\bar{n}}}{8\Omega_{2g}N^{1/2}} \lambda^2_{\hat{S}_\alpha} $ \\
\hline
motional dephasing & Eq.~(\ref{eq:dephasing_fid}) & $\mathcal{I}^{2\eta}=\frac{\pi \eta}{16\Omega_{2g} N^{1/2}}\Big[\Big(2\bar{n}_{a}+1\Big)\lambda^2_{\hat{S}_\alpha} + \frac{3}{16N}\lambda^2_{\hat{S}_\alpha^2}\Big]$  \\
 \hline
\end{tabular}
\caption{Summary of infidelities for $N$-loop two-qubit gates. See text for variable definitions.}
\end{table*}

\subsubsection{Motional dephasing}\label{sec:dephasing}

In Sec.~\ref{sec:static_motion}, we derived the infidelity of a gate due to a static shift in motional frequency. In this section, we derive the effects of non-static shifts in the limit of vanishingly small coherence times. We begin by adding an error term:
\begin{eqnarray}\label{eq:motional_deph_ham}
\hat{H}_{2\omega}^{g_{d}}(t) = \hbar g_{d}\cos(\omega t)\hat{a}^{\dagger}\hat{a},
\end{eqnarray}
which adds a sinusoidally oscillating frequency shift to the motional mode, giving $\hat{H}_{2t}=\hat{H}_{2g}+\hat{H}_{2\omega}^{g_{d}}$ for the total system Hamiltonian. In this section, we first transform $\hat{H}_{2t}$ into the interaction picture with respect to $\hat{H}_{2g}$, which gives: 
\begin{eqnarray}\label{eq:markovian_interaction_pic}
\hat{H}_{2I}^{g_{d}} = \hbar g_{d}\cos(\omega t)[\hat{a}^{\dagger}+\hat{S}_{\alpha}\gamma^{*}][\hat{a}+\hat{S}_{\alpha}\gamma],
\end{eqnarray}
where $\gamma$ is here defined by Eq.~(\ref{eq:gate_gamma}). We now determine $\hat{U}_{2I}^{g_{d}}$ using Eq.~(\ref{eq:time_dep_orig}), leaving the integrals unevaluated for now. Since this results in a factored $\hat{U}_{2t}^{g_{d}}=\hat{U}_{2g}\hat{U}_{2I}^{g_{d}}$, we can apply Eq.~(\ref{eq:fn_just_error}). Dropping all terms higher-order than $\propto  g_{d}^{2}$ gives:
\begin{eqnarray}\label{eq:mot_deph_omega_fid}
\mathcal{F}_{n_{a},\omega}^{g_{d}} &=& 1 -2 g_{d}^{2}\!\int^{t_{g}}_{0}\!\!\int^{t^{\prime}}_{0}dt^{\prime}dt^{\prime\prime}\cos(\omega t^{\prime})\cos(\omega t^{\prime\prime})\Big[n_{a}^{2} \nonumber \\
&&+\braket{\hat{S}_{\alpha}^{2}}\Big\{n_{a}\big(|\gamma(t^{\prime})|^{2} + |\gamma(t^{\prime\prime})|^{2}\big) + n_{a}\gamma(t^{\prime})\gamma^{*}(t^{\prime\prime}) \nonumber \\
&&+ (n_{a}+1)\gamma^{*}(t^{\prime})\gamma(t^{\prime\prime})\Big\} + \braket{\hat{S}_{\alpha}^{4}}|\gamma(t^{\prime})|^{2}|\gamma(t^{\prime\prime})|^{2} \Big] \nonumber \\
&&+ g_{d}^{2}\int^{t_{g}}_{0}\int^{t_{g}}_{0}dt^{\prime}dt^{\prime\prime}\cos(\omega t^{\prime})\cos(\omega t^{\prime\prime})\Big[n_{a}^{2} \nonumber \\
&&+\braket{\hat{S}_{\alpha}^{2}}\!n_{a}\big(|\gamma(t^{\prime})|^{2} + |\gamma(t^{\prime\prime})|^{2}\big)\! +\!\nonumber \\&& 
\braket{\hat{S}_{\alpha}}^{2}\Big\{ n_{a}\gamma^{*}(t^{\prime})\gamma(t^{\prime\prime})+(n_{a}+1)\gamma(t^{\prime})\gamma^{*}(t^{\prime\prime})\Big\}\nonumber\\
&&+ \braket{\hat{S}_{\alpha}^{2}}^{2}|\gamma(t^{\prime})|^{2}|\gamma(t^{\prime\prime})|^{2}\Big].
\end{eqnarray}
We can plug this equation into Eq.~(\ref{eq:avg_over_omega}) to obtain $\mathcal{F}^{g_{d}}_{n_{a}}$. Upon doing this, we are left with a sum of triple integrals, each of which is proportional to:
\begin{eqnarray}\label{eq:integrals_mot_deph}
\zeta &=& \!\int^{\infty}_{0}\!\!\int^{t_{g}}_{0}\!\int^{t_{s}}_{0}\! d\omega dt^{\prime}dt^{\prime\prime}\frac{S_{\omega}}{2}\Big\{\!\cos(\omega[t^{\prime\prime}\!+\!t^{\prime}]) \!+\!\cos(\omega[t^{\prime\prime}\!-\!t^{\prime}])\Big\}, \nonumber \\
\end{eqnarray}
where $t_{s}\in\{t_{g},t^{\prime}\}$. In order to evaluate Eq.~(\ref{eq:integrals_mot_deph}), we perform the following manipulations:
\begin{eqnarray}
\zeta &=& \!\int^{\infty}_{-\infty}\!\int^{t_{g}}_{0}\!\int^{t_{s}}_{0}d\omega dt^{\prime}dt^{\prime\prime}\frac{S_{\omega}}{4}\Big\{\!\cos(\omega[t^{\prime\prime}\! +\! t^{\prime}]) \!+\!\cos(\omega[t^{\prime\prime}\!-\! t^{\prime}])\!\Big\} \nonumber \\
&\simeq& \frac{S_{0}}{4}\!\int^{\infty}_{-\infty}\!\int^{t_{g}}_{0}\!\int^{t_{s}}_{0}\! d\omega  dt^{\prime} dt^{\prime\prime}\!\Big\{\!\cos(\omega[t^{\prime\prime}\!+\! t^{\prime}]) \!+\!\cos(\omega[t^{\prime\prime}\!-\! t^{\prime}])\Big\} \nonumber \\
&=& \frac{\pi S_{0}}{2}\int^{t_{g}}_{0}\int^{t_{s}}_{0} dt^{\prime}dt^{\prime\prime}\Big\{\delta(t^{\prime\prime}+t^{\prime}) + \delta(t^{\prime\prime}-t^{\prime})\Big\},
\end{eqnarray}
where, in the second line, we assumed a white noise bath of $\omega$ and pulled $S_{\omega}\simeq S_{0}$ outside of the integral. We can now straightforwardly evaluate the integrals in Eq.~(\ref{eq:mot_deph_omega_fid}), which gives:
\begin{eqnarray}
\mathcal{F}_{n_{a}}^{g_{d}} \simeq 1 - \pi g_{d}^{2}S_{0}\Big(\frac{\Omega_{2g}^{2}t_{g}}{\Delta^{2}} \Big)\Big([2n_{a}+1]\lambda^{2}_{\hat{S}_{\alpha}} + \frac{3\Omega_{2g}^{2}}{\Delta^{2}}\lambda^{2}_{\hat{S}_{\alpha}^{2}} \Big), \nonumber \\
\end{eqnarray}
where, plugging in $t_{g}=2\pi N/\Delta$ and $\Delta = 4\Omega_{2g}N^{1/2}$ gives a final infidelity of:
\begin{eqnarray}\label{eq:dephasing_fid}
\mathcal{I}_{n_{a}}^{g_{d}} &=& \frac{\pi \eta}{16\Omega_{2g} N^{1/2}}\Big([2n_{a}+1]\lambda^{2}_{\hat{S}_{\alpha}} + \frac{3}{16N}\lambda^{2}_{\hat{S}_{\alpha}^{2}}\Big),
\end{eqnarray}
where $\eta\equiv\pi g_{d}^{2}S_{0}/2$ (see appendix). Note that $\eta=2/\tau$, where $\tau$ corresponds to the decay time of the coherence between two neighboring Fock states. In Fig.~\ref{fig:coherent_two_qubit_markovian}(b), we compare Eq.~(\ref{eq:dephasing_fid}) to the direct numerical integration of the motional dephasing master equation (see appendix), showing the two calculations converge when $\eta/\Omega_{2g}\ll 1$ for various initial states of the qubits and motion. 

\section{Conclusion}

In this work, we derived formulae describing how several motional error sources in trapped ions and trapped electrons affect the fidelity of one- and two-qubit gates. The effect of these error sources on infidelities are typically calculated numerically when determining an individual experiment's error budget. Therefore, this work serves to both expedite the creation of error budgets, and provide physicists with a deeper understanding of how these gate infidelities depend on the parameters of their experiments: temperature, initial qubit state, the number of loops traversed in phase-space, and so on. Finally, we compare all our analytic derivations to their respective numerical simulations, showing they converge for high-fidelity gates.

\section*{Acknowledgements}
R.T.S., Q.Y. and K.M.B. acknowledge support from the Lawrence Livermore National Laboratory (LLNL) Laboratory Directed Research and Development (LDRD) program under Grant No. 21-FS-008. Q.Y. and H.H. acknowledge support from AFOSR through grant FA9550-20-1-0162, the NSF QLCI program through grant number OMA-2016245. K.M.B's contributions to this work were performed under the auspices of the U. S. Department of Energy by Lawrence Livermore National Laboratory under Contract No. DE-AC52-07NA27344. LLNL-JRNL-828517

\section*{Appendix}

\subsection{Static Qubit frequency shifts}

If the qubit frequency is erroneously shifted from its desired value during the course of a one-qubit gate, the Hamiltonian is given by:
\begin{eqnarray}\label{eq:one_qubit_shifts}
\hat{H}_{1t} &=& \hat{H}_{1g} + \hat{H}_{1e}^{q} \nonumber \\
&=& \hbar\Omega_{1g}\hat{\sigma}_{x} + \hbar\delta\hat{\sigma}_{z},
\end{eqnarray}
where we have assumed that the gate operation is polarized in the $\hat{x}$-direction. Equation~(\ref{eq:one_qubit_shifts}) acts as a rotation about the $\hat{n}$ axis, where:
\begin{eqnarray}
    \hat{n} \equiv \frac{\Omega_{1g}\hat{x} + \delta \hat{z}}{\sqrt{\Omega_{1g}^{2} + \delta^{2}}},
\end{eqnarray}
at an angular frequency:
\begin{eqnarray}
    \Omega^{\prime}_{1g}\equiv \sqrt{\Omega^{2}_{1g} + \delta^{2}}.
\end{eqnarray}
This gives a time evolution operator:
\begin{eqnarray}\label{eq:u_one_shift_err}
    \hat{U}_{t}(t) &=& e^{-i\Omega^{\prime}_{1g}t(\hat{n} \cdot \vec{\sigma})} \nonumber \\
        &=& \hat{I}\cos(\Omega^{\prime}_{1g}t) - i(\hat{n}\cdot\vec{\sigma})\sin(\Omega^{\prime}_{1g}t).
\end{eqnarray}
This is the exact solution for the time evolution of Eq.~(\ref{eq:one_qubit_shifts}) \cite{agarwal_2012}. In this work, however, we concerned with calculating the the infidelity $\mathcal{I}^{1q}$ due to small values of $\delta$. We, therefore, relate $\hat{U}_{t}$ to $\hat{U}_{1g}$ by expanding $\Omega^{\prime}_{1g}$ and $\hat{n}$ about the point $\delta = 0$, which gives:
\begin{eqnarray}
    \Omega^{\prime}_{1g} &\simeq& \Omega_{1g} + \frac{\delta^{2}}{2\Omega_{1g}} \nonumber \\
    \hat{n} &\simeq &\Big(1 - \frac{\delta^{2}}{2\Omega_{1g}^2} \Big) \hat{x} + \Big(\frac{\delta}{\Omega_{1g}}\Big)\hat{z}.
\end{eqnarray}
After expanding these two variables, we can also expand the sinusoidal functions that appear in Eq.~(\ref{eq:u_one_shift_err}):
\begin{eqnarray}
    \cos(\Omega^{\prime}_{1g}t) &\simeq & \cos(\Omega_{1g} t) - \Big(\frac{\delta^{2}t}{2\Omega_{1g}} \Big)\sin(\Omega_{1g} t) \nonumber \\
    \sin(\Omega^{\prime}_{1g} t) &\simeq & \sin(\Omega_{1g} t) + \Big(\frac{\delta^{2}t}{2\Omega_{1g}} \Big)\cos(\Omega_{1g} t).
\end{eqnarray}
Plugging these expansions into Eq.~(\ref{eq:u_one_shift_err}) gives:
\begin{eqnarray}
    \hat{U}_{1g}\Big\{1 \!-\! i\Big(\frac{\delta^{2}t}{2\Omega_{1g}} \Big)\hat{\sigma}_{x} \Big\} \!+\! i\sin(\Omega_{1g}t)\Big\{\Big(\frac{\delta^{2}}{2\Omega^{2}_{1g}}\Big)\hat{\sigma}_{x} \!-\! \Big(\frac{\delta}{\Omega_{1g}}\Big)\hat{\sigma}_{z}\Big\}. \nonumber \\
\end{eqnarray}
We can then plug this equation into Eq.~(\ref{eq:initial_fn}), which gives:
\begin{eqnarray}
\mathcal{F} &=& \Big|\bra{\psi(0)}\Big\{1-i\Big(\frac{\delta^{2}t}{2\Omega_{1g}}\Big)\hat{\sigma}_{x} \Big\} + i \sin(\Omega_{1g} t)\hat{U}_{1g}^{ \dagger}\Big\{\Big(\frac{\delta^{2}}{2\Omega_{1g}^2}\Big)\hat{\sigma}_{x} \nonumber \\
&&- \Big(\frac{\delta}{\Omega_{1g}}\Big)\hat{\sigma}_{z} \Big\}\ket{\psi(0)}\Big|^{2}.
\end{eqnarray}
Only keeping terms up to $\mathcal{O}([\delta/\Omega_{1g}]^{2})$, we get:
\begin{eqnarray}
\mathcal{F} &\simeq & 1 - \Big(\frac{\delta}{\Omega_{1g}}\Big)^{2}\Big\{\sin^{2}(\Omega_{1g}t) \nonumber \\
&&- \Big[\cos(\Omega_{1g}t)\sin(\Omega_{1g}t)\braket{\hat{\sigma}_{z}} + \sin^{2}(\Omega_{1g}t)\braket{\hat{\sigma}_{y}}\Big]^{2}\Big\}. \nonumber \\
\end{eqnarray}
If we assume that $\ket{\psi(0)} = \ket{0}$, i.e. $\braket{\sigma_{z}}= - 1$ and $\braket{\sigma_{y}} = 0$, this gives an infidelity of:
\begin{eqnarray}
\mathcal{I}^{1q}\simeq \Big(\frac{\delta^{2}}{\Omega_{1g}^{2}}\Big)\sin^{4}(\Omega_{1g}t).
\end{eqnarray}
If we average $\ket{\psi(0)}$ over the Bloch sphere, this gives:
\begin{eqnarray}
\label{eq:1Q static frequency shift infidelity}
\mathcal{I}^{1q}\simeq \Big(\frac{2\delta^{2}}{3\Omega_{1g}^{2}} \Big)\sin^{2}(\Omega_{1g} t).
\end{eqnarray}

\subsection{Static motional frequency shifts with Walsh sequences}

In Sec.~\ref{sec:static_motion}, we showed that a geometric phase gate traversing a single loop in phase-space produces an error operator of:
\begin{eqnarray}\label{eq:error_static_mot_one_loop}
    \hat{U}_{2e}^{\delta,0} \equiv \exp\Big(-\frac{i\Omega_{2g}\delta t_{l}}{\Delta}\hat{S}_{\alpha}[\hat{a} + \hat{a}^{\dagger}] - \frac{2i\Omega_{2g}^{2}\delta t_{l}}{\Delta^{2}}\hat{S}_{\alpha}^{2}\Big), \nonumber \\
\end{eqnarray}
where we have replaced the gate time $t_{g}$ in Eq.~(\ref{eq:error_static_mot_shift}) with a single loop time $t_{l}=2\pi/\Delta$. The error described by $\hat{U}_{2e}^{\delta,0}$ comprises a spin-dependent ($\propto \hat{S}_{\alpha}$) displacement operator, representing the residual spin-motion entanglement, and a $\propto \hat{S}_{\alpha}^{2}$ operator, representing the error in the geometric phase. Reference~\cite{hayes_2012} showed that administering pi-pulses such that $\hat{H}\rightarrow -\hat{H}$ can suppress the former of these. To see how this works, we first must understand that the $k^{th}$-order Walsh sequence $W(2^{k}-1,x)$ is simply two concatenated $(k-1)^{th}$-order Walsh sequences, where $\hat{H}\rightarrow -\hat{H}$ for the latter of the two. If the error operator for the $(k-1)^{th}$-order Walsh function takes the form:
\begin{eqnarray}
\hat{U}_{2e}^{\delta,k-1} = \exp\Big(-i\hat{S}_{\alpha}\varepsilon\Big[\gamma\hat{a}^{\dagger} +\gamma^{*} \hat{a}\Big] \Big),
\end{eqnarray}
where $\varepsilon$ is an arbitrary constant, and:
\begin{eqnarray}
\gamma(t) &=& \int^{t}_{t_{0}}dt^{\prime}e^{i(\Delta + \delta)t^{\prime}} \nonumber \\
&=& e^{i(\Delta +\delta)t_{0}}\int^{t-t_{0}}_{0}dt^{\prime}e^{i(\Delta +\delta)t^{\prime}} \nonumber \\
&=& e^{i\delta t_{0}}\int^{t-t_{0}}_{0}dt^{\prime}e^{i(\Delta +\delta)t^{\prime}},
\end{eqnarray}
where, in the third-line, we have assumed that that $t_{0}$ is an integer multiple of $2\pi/\Delta$. If $t_{k-1}$ is the time it takes a $(k-1)^{th}$-order Walsh sequence to complete, $\gamma(t)$ for the second of the two concatenated Walsh sequences is:
\begin{eqnarray}
\gamma(2t_{k-1}) &=& \int^{2t_{k-1}}_{t_{k-1}}dt^{\prime}e^{i(\Delta + \delta)t^{\prime}} \nonumber \\
&=& e^{i\delta t_{k-1}}\int^{t_{k-1}}_{0}dt^{\prime}e^{i(\Delta +\delta)t^{\prime}},
\end{eqnarray}
meaning that $\gamma$ for the second sequence will be identical to the first, up to a phase $e^{i\delta t_{k-1}}$. Keeping this in mind, we get:
\begin{eqnarray}
\hat{U}^{\delta,k}_{2e} &=& \!\exp\!\Big(\! i\hat{S}_{\alpha}\varepsilon\! \Big[e^{i\delta t_{k-1}}\gamma\hat{a}^{\dagger}\!\! + \!e^{-i\delta t_{k-1}}\gamma^{*} \hat{a}\Big] \Big)\!\exp\!\Big(\!\!-\! i\hat{S}_{\alpha}\varepsilon\Big[\gamma\hat{a}^{\dagger} \!+\!\gamma^{*} \hat{a}\Big] \Big) \nonumber \\
&\simeq &\exp\Big(\!\!-i\hat{S}_{\alpha}\varepsilon(i\delta t_{k-1})\Big[-\gamma\hat{a}^{\dagger}+\gamma^{*}\hat{a} \Big]\Big),
\end{eqnarray}
showing that increasing $k$ by one reduces the argument of $\hat{U}^{\delta,k-1}_{2e}$ by a factor of $i\delta t_{k-1}$. A $W(1,x)$ Walsh sequence is a two loop gate where $\hat{H}\rightarrow -\hat{H}$ after the first loop at $t_{k-1}=t_{l}$. This gives:
\begin{eqnarray}
\hat{U}^{\delta, 1}_{2e} &=& \exp\!\Big(\!-\frac{\Omega_{2g}(i\delta t_{l})^{2}}{\Delta}\hat{S}_{\alpha}[\hat{a}-\hat{a}^{\dagger}]- \frac{2i\Omega_{2g}^{2}\delta t_{g}}{\Delta^{2}}\hat{S}_{\alpha}^{2}\Big), \nonumber
\end{eqnarray}
where $t_{g}=2\pi N/\Delta$, where $N=2^{k}$ is the number of loops in phase space. This process can be repeated for a $W(3,x)$ sequence, which is just two concatenated $W(1,x)$ sequences, such that $\hat{H}\rightarrow -\hat{H}$ after the first sequence at $t_{k-1}=2t_{l}$:
\begin{eqnarray}
    \hat{U}^{\delta,2}_{2e} \!= \exp\!\Big(\! -\frac{\Omega_{2g} (i\delta t_{l})^{3}}{\Delta}(1\cdot 2)\hat{S}_{\alpha}[\hat{a}+\hat{a}^{\dagger}] - \frac{2i\Omega_{2g}^{2}\delta t_{g}}{\Delta^{2}}\hat{S}_{\alpha}^{2}\Big). \nonumber \\
\end{eqnarray}
This pattern can be repeated to give the error propagator for a general $k^{th}$-order Walsh sequence:
\begin{eqnarray}
\hat{U}^{\delta,k}_{2e} &=& \exp\!\Big(\!\!-\frac{\Omega_{2g}(i\delta t_{l})^{k+1}2^{\frac{k(k-1)}{2}}}{\Delta}\hat{S}_{\alpha}[\hat{a}\!+\!(-1)^{k}\hat{a}^{\dagger}] \nonumber \\
&&~~~~~~-\frac{2i\Omega_{2g}^{2}\delta t_{g}}{\Delta^{2}}\hat{S}_{\alpha}^{2}\Big).
\end{eqnarray}
This error operator can be plugged into Eq.~(\ref{eq:fn_just_error}) and Taylor expanded to find the leading-order corrections to the gate infidelity:
\begin{eqnarray}
\mathcal{I}^{\delta,k}_{n_{a}} \!&=&\! 2^{k(k-1)}\frac{\Omega_{2g}^{2}(\delta t_{l})^{2(k+1)}}{\Delta^{2}} (2n_{a}+1)\lambda^{2}_{\hat{S}_{\alpha}} \!\!+\!\! \frac{4\Omega_{2g}^{4}\delta^{2}t_{g}^{2}}{\Delta^{4}}\lambda_{\hat{S}_{\alpha}^{2}}, \nonumber \\
\end{eqnarray}
which we can simplify by substituting $\Delta = 4\Omega_{2g}N^{1/2}$, $t_{l}=2\pi/\Delta$, $t_{g}=2\pi N/\Delta$, and $N=2^{k}$. Upon averaging over $\mathcal{P}_{n_{a}}$, we obtain the infidelity of a two-qubit gate undergoing a $W(2^{k}-1,x)$ Walsh sequence, for an arbitrary initial state of the qubits' and mixed-state of the motion:
\begin{eqnarray}
\mathcal{I}^{\delta,k}\! &=& \!\Big(\frac{\pi\delta}{\Omega_{2g}} \Big)^{2(k+1)}\!2^{-(5k+6)}(2\bar{n}_{a}\!+\!1)\lambda^{2}_{\hat{S}_{\alpha}}\!\!+\!\frac{\pi^{2}\delta^{2}}{\Omega_{2g}^{2}} 2^{-(k+8)}\lambda^{2}_{\hat{S}_{\alpha}^{2}}.\nonumber \\
\end{eqnarray}

\subsection{Markovian master equations}

Equation~(\ref{eq:lindblad}) contains three triple integrals:
\begin{eqnarray}\label{eq:linblad_integrals}
\Upsilon_{i} &=& \frac{1}{\hbar^{2}\delta t}\!\int^{\infty}_{0}\!\!\!\int^{t+\delta t}_{t}\!\!\!\int^{t+\delta t}_{t}\!\!\!d\omega dt^{\prime}dt^{\prime\prime}S_{\omega}\hat{H}_{2\omega}({t^{\prime}})\hat{\rho}(t)\hat{H}_{2\omega}(t^{\prime\prime}) \nonumber \\
\Upsilon_{ii} &=& \frac{1}{\hbar^{2}\delta t}\!\int^{\infty}_{0}\!\!\int^{t+\delta t}_{t}\!\!\int^{t^{\prime}}_{t}\!\!\!d\omega dt^{\prime}dt^{\prime\prime}S_{\omega}\hat{\rho}(t)\hat{H}_{2\omega}({t^{\prime\prime}})\hat{H}_{2\omega}(t^{\prime}) \nonumber \\
\Upsilon_{iii} &=& \frac{1}{\hbar^{2}\delta t}\!\int^{\infty}_{0}\!\!\int^{t+\delta}_{t}\!\!\int^{t^{\prime}}_{t}\!\!\!d\omega dt^{\prime}dt^{\prime\prime}S_{\omega}\hat{H}_{2\omega}({t^{\prime}})\hat{H}_{2\omega}(t^{\prime\prime})\hat{\rho}(t). \nonumber \\
\end{eqnarray}
Once we determine $\hat{H}_{2\omega}(t)$, we can evaluate these integrals using the similar approximations to that in the text.

\subsubsection*{Heating master equation}

We begin with Eq.~(\ref{eq:interaction_heat}), representing a stray electric field at frequency $\omega$, taken in the rotating frame with respect to the frequency of the trap $\omega_{a}$, and makes the rotating wave approximation:
\begin{eqnarray}
\hat{H}^{g_{h}\prime}_{2\omega}(t)\simeq \hbar g_{h}\Big(\hat{a}^{\dagger}e^{i[\omega_{a}-\omega]t} + \hat{a}e^{-i[\omega_{a}-\omega]t} \Big). 
\end{eqnarray}
Plugging this into $\Upsilon_{i}$ in Eq.~(\ref{eq:linblad_integrals}) gives:
\begin{eqnarray}
\Upsilon_{i} &=& \frac{g_{h}^{2}}{\delta t}\!\!\int^{\infty}_{0}\!\!\int^{t+\delta t}_{t} \!\!\!\int^{t+\delta t}_{t}d\omega dt^{\prime}dt^{\prime\prime}S_{\omega}\Big\{ \hat{a}^{\dagger}\hat{\rho}(t)\hat{a}^{\dagger}e^{i\omega^{\prime}(t^{\prime\prime}+t^{\prime})} \nonumber \\
&&+ ~\hat{a}\hat{\rho}(t)\hat{a}e^{-i\omega^{\prime}(t^{\prime\prime}\!+t^{\prime})} \!+ \hat{a}^{\dagger}\hat{\rho}(t)\hat{a}e^{-i\omega^{\prime}(t^{\prime\prime}\!-t^{\prime})} \nonumber \\
&&+ ~\hat{a}\hat{\rho}(t)\hat{a}^{\dagger}e^{i\omega^{\prime}(t^{\prime\prime}\!-t^{\prime})}\!\Big\},
\end{eqnarray}
where $\omega^{\prime}\equiv \omega_{a}-\omega$. We can evaluate this integral by first replacing $S_{\omega}$ with a constant $S_{\omega_{a}}$:
\begin{eqnarray}
\Upsilon_{i} &\simeq & \frac{g_{h}^{2} S_{\omega_{a}}}{\delta t}\!\!\int^{\infty}_{0}\!\!\int^{t+\delta t}_{t}\!\!\int^{t+\delta t}_{t}\!\!\!d\omega dt^{\prime}dt^{\prime\prime}\Big\{\hat{a}^{\dagger}\hat{\rho}(t)\hat{a}^{\dagger}e^{i\omega^{\prime}(t^{\prime\prime}+t^{\prime})} \nonumber \\
&& + \hat{a}\hat{\rho}(t)\hat{a}e^{-i\omega^{\prime}(t^{\prime\prime}\!+t^{\prime})}\! \!+\! \hat{a}^{\dagger}\hat{\rho}(t)\hat{a}e^{-i\omega^{\prime}(t^{\prime\prime}\!-t^{\prime})}\!\! +\! \hat{a}\hat{\rho}(t)\hat{a}^{\dagger}e^{i\omega^{\prime}(t^{\prime\prime}-t^{\prime})}\Big\} \nonumber \\
&\simeq & \frac{g_{h}^{2} S_{\omega_{a}}}{2\delta t}\int^{\infty}_{-\infty}\!\!\int^{t+\delta t}_{t}\int^{t+\delta t}_{t}\!\!\!d\omega dt^{\prime}dt^{\prime\prime}\Big\{\hat{a}^{\dagger}\hat{\rho}(t)\hat{a}^{\dagger}e^{i\omega^{\prime}(t^{\prime\prime}+t^{\prime})} \nonumber \\
&& + \hat{a}\hat{\rho}(t)\hat{a}e^{-i\omega^{\prime}(t^{\prime\prime}+t^{\prime})}\!\! +\! \hat{a}^{\dagger}\hat{\rho}(t)\hat{a}e^{-i\omega^{\prime}(t^{\prime\prime}\!-t^{\prime})} \!\!+ \hat{a}\hat{\rho}(t)\hat{a}^{\dagger}e^{i\omega^{\prime}(t^{\prime\prime}\!-t^{\prime})}\Big\} \nonumber \\
&= & \frac{\pi S_{\omega_{a}}g_{h}^{2}}{\delta t}\int^{t+\delta t}_{t}\int^{t+\delta t}_{t}dt^{\prime}dt^{\prime\prime}\Big\{\delta(t^{\prime\prime}+t^{\prime})[\hat{a}^{\dagger}\hat{\rho}(t)\hat{a}^{\dagger} + \hat{a}\hat{\rho}(t)\hat{a}] \nonumber \\
&&+\delta(t^{\prime\prime}-t^{\prime})[\hat{a}^{\dagger}\hat{\rho}(t)\hat{a} + \hat{a}\hat{\rho}(t)\hat{a}^{\dagger}] \Big\} \nonumber \\
&= & \pi S_{\omega_{a}}g_{h}^{2}[\hat{a}^{\dagger}\hat{\rho}(t)\hat{a}+\hat{a}\hat{\rho}(t)\hat{a}^{\dagger}] \nonumber \\
&=& \dot{\bar{n}}[\hat{a}^{\dagger}\hat{\rho}(t)\hat{a}+\hat{a}\hat{\rho}(t)\hat{a}^{\dagger}],
\end{eqnarray}
where, in the last line, we have introduced the heating rate $\dot{\bar{n}}\equiv \pi S_{\omega_{a}}g_{h}^{2}$. Keeping in mind the added factor of $1/2$ that comes from changing the limits of integration from $t+\delta t$ to $t^{\prime}$ in the integral over $dt^{\prime\prime}$, we can follow this prescription to evaluate $\Upsilon_{ii}$ and $\Upsilon_{iii}$. This gives a final master equation for Markovian heating of:
\begin{eqnarray}
\dot{\hat{\rho}}(t) &=& -\frac{i}{\hbar}\Big[\hat{H}_{2g}(t),\hat{\rho}(t) \Big] + \dot{\bar{n}}\Big\{\hat{a}^{\dagger}\hat{\rho}(t)\hat{a} + \hat{a}\hat{\rho}(t)\hat{a}^{\dagger} \nonumber \\
&&- \frac{1}{2}\Big(\hat{a}^{\dagger}\hat{a} + \hat{a}\hat{a}^{\dagger} \Big)\hat{\rho}(t) - \frac{1}{2}\hat{\rho}(t)\Big(\hat{a}^{\dagger}\hat{a}+\hat{a}\hat{a}^{\dagger} \Big)\Big\} \nonumber \\
\end{eqnarray}

\subsubsection*{Motional dephasing master equation}
We begin with Eq.~(\ref{eq:motional_deph_ham}), representing fluctuations of the motional frequency of the trap at frequency $\omega$ as:
\begin{eqnarray}
\hat{H}_{2\omega}^{g_{d}}(t) = \hbar g_{d}\cos(\omega t)\hat{a}^{\dagger}\hat{a}.
\end{eqnarray}
Plugging this into $\Upsilon_{i}$ in Eq.~(\ref{eq:linblad_integrals}) gives:
\begin{eqnarray}
\Upsilon_{i} &=& \frac{g_{d}^{2}}{\delta t}\!\int^{\infty}_{0}\!\!\!\int^{t+\delta t}_{t}\!\!\!\!\int^{t+\delta t}_{t}\!\!\!\!\!\!d\omega dt^{\prime}dt^{\prime\prime}\!S_{\omega}\!\cos(\omega t^{\prime})\!\cos(\omega t^{\prime\prime})\hat{a}^{\dagger}\hat{a}\hat{\rho}(t)\hat{a}^{\dagger}\hat{a} \nonumber \\
&=& \frac{g_{d}^{2}}{2\delta t}\int^{\infty}_{0}\int^{t+\delta t}_{t}\int^{t+\delta t}_{t}d\omega dt^{\prime}dt^{\prime\prime}S_{\omega}\Big\{\cos(\omega[t^{\prime\prime}+t^{\prime}]) \nonumber \\
&&+ \cos(\omega[t^{\prime\prime}-t^{\prime}]) \Big\}\hat{a}^{\dagger}\hat{a}\hat{\rho}(t)\hat{a}^{\dagger}\hat{a}.
\end{eqnarray}
We can, again, make the approximation of replacing the $S_{\omega}$ term with a constant $S_{0}$, and change the limits of the integral over $\omega$ to obtain:
\begin{eqnarray}
\Upsilon_{i} &\simeq& \frac{g_{d}^{2}S_{0}}{4\delta t}\int^{\infty}_{-\infty}\int^{t+\delta t}_{t}\int^{t+\delta t}_{t}d\omega dt^{\prime}dt^{\prime\prime}\Big\{\cos(\omega[t^{\prime\prime}+t^{\prime}]) \nonumber \\
&&+ \cos(\omega[t^{\prime\prime}-t^{\prime}]) \Big\}\hat{a}^{\dagger}\hat{a}\hat{\rho}(t)\hat{a}^{\dagger}\hat{a} \nonumber \\
&=& \frac{\pi g_{d}^{2}S_{0}}{2\delta t}\int^{t+\delta t}_{t}\!\!\!\!\int^{t+\delta t}_{t}\!\!\!\!\!\!dt^{\prime}dt^{\prime\prime}\Big\{\!\delta(t^{\prime\prime}\!+\!t^{\prime}) + \delta(t^{\prime\prime}-t^{\prime})\! \Big\}\hat{a}^{\dagger}\hat{a}\hat{\rho}(t)\hat{a}^{\dagger}\hat{a} \nonumber \\
&=& \frac{\pi g_{d}^{2}S_{0}}{2}\hat{a}^{\dagger}\hat{a}\hat{\rho}(t)\hat{a}^{\dagger}\hat{a} \nonumber \\
&=& \eta\hat{a}^{\dagger}\hat{a}\hat{\rho}(t)\hat{a}^{\dagger}\hat{a},
\end{eqnarray}
where, in the last line, we have introduced the motional dephasing rate $\eta\equiv \pi g_{d}^{2}S_{0}/2$.
Again, keeping in mind the limits of integration over $dt^{\prime\prime}$, $\Upsilon_{ii}$ and $\Upsilon_{iii}$ can be evaluated in the same manner. This gives a final equaiton for Markovian motional dephasing:
\begin{eqnarray}
\dot{\hat{\rho}}(t) &=& -\frac{i}{\hbar}\Big[\hat{H}_{2g}(t),\hat{\rho}(t) \Big] \nonumber \\
&&+ \eta\Big\{\hat{a}^{\dagger}\hat{a}\hat{\rho}(t)\hat{a}^{\dagger}\hat{a} - \frac{1}{2}[\hat{a}^{\dagger}\hat{a}]^{2}\hat{\rho}(t) - \frac{1}{2}\hat{\rho}(t)[\hat{a}^{\dagger}\hat{a}]^{2} \Big\}. \nonumber \\
\end{eqnarray}

\bibliography{biblio}

\end{document}